\documentclass[journal]{IEEEtran}
\ifCLASSINFOpdf
\else
\fi
\usepackage{dblfloatfix}
\usepackage{amsmath}
\usepackage{graphicx}
\usepackage[numbers,sort&compress]{natbib}
\usepackage{multicol}
\usepackage[colorlinks,citecolor=blue,urlcolor=blue,pdftitle={Octopus},pdfauthor={Mari},pdfsubject={LaTeX}]{hyperref}
\usepackage{multirow}
\usepackage{array}
\usepackage{hhline}
\usepackage{rotating}
\usepackage[shortlabels]{enumitem}
\usepackage[font=small,labelfont=small]{caption}

\usepackage[10pt]{moresize}

\newcolumntype{P}[1]{>{\centering\arraybackslash}p{#1}}
\newcolumntype{M}[1]{>{\centering\arraybackslash}m{#1}}
\newcolumntype{R}[1]{>{\raggedleft\arraybackslash}p{#1}}
\newcommand{\norm}[1]{\left\lVert#1\right\rVert}

\usepackage{tikz}
\usepackage{textcomp}

\newcommand\copyrighttext{%
	\footnotesize \textcopyright 2021 IEEE.  Personal use of this material is permitted.  Permission from IEEE must be obtained for all other uses, in any current or future media, including reprinting/republishing this material for advertising or promotional purposes, creating new collective works, for resale or redistribution to servers or lists, or reuse of any copyrighted component of this work in other works.
	DOI: \href{https://doi.org/10.1109/TIFS.2021.3067998}{10.1109/TIFS.2021.3067998}}
\newcommand\copyrightnotice{%
	\begin{tikzpicture}[remember picture,overlay]
		\node[anchor=south,yshift=10pt] at (current page.south) {\fbox{\parbox{\dimexpr\textwidth-\fboxsep-\fboxrule\relax}{\copyrighttext}}};
	\end{tikzpicture}%
}

\usepackage[resetlabels]{multibib}
\newcites{latex}{References}
\begin{document}
%
\title{Evidence of Task-Independent Person-Specific Signatures in EEG using Subspace Techniques}
%
%
\author{Mari~Ganesh~Kumar,~\IEEEmembership{Student Member,~IEEE,}
        Shrikanth~Narayanan,~\IEEEmembership{Fellow,~IEEE,}
        Mriganka~Sur,~\IEEEmembership{Member,~IEEE}
        and~Hema~A~Murthy,~\IEEEmembership{Senior Member,~IEEE}
\thanks{M. G. Kumar and H. A. Murthy are with the Department
of Computer Science and Engineering, Indian Institute of Technology Madras, Chennai, India e-mail: \{mari, hema\}@cse.iitm.ac.in}
\thanks{S. Narayanan is with the Viterbi School of Engineering, University of Southern California, USA e-mail: shri@ee.usc.edu}
\thanks{M. Sur is with the Department of Brain and Cognitive Sciences, Massachusetts Institute of Technology, USA e-mail: msur@mit.edu}}
\maketitle

\copyrightnotice

\begin{abstract}
Electroencephalography (EEG) signals are promising as alternatives to other biometrics  owing to their protection against spoofing. Previous studies have focused on capturing individual variability by analyzing task/condition-specific EEG. This work attempts to model biometric signatures independent of task/condition by normalizing the associated variance. Toward this goal, the paper extends ideas from subspace-based text-independent speaker recognition and proposes novel modifications for modeling multi-channel EEG data. The proposed techniques assume that biometric information is present in the entire EEG signal and accumulate statistics across time in a high dimensional space. These high dimensional statistics are then projected to a lower dimensional space where the biometric information is preserved. The lower dimensional embeddings obtained using the proposed approach are shown to be task-independent. The best subspace system identifies individuals with accuracies of $86.4\%$ and $35.9\%$ on datasets with $30$ and $920$ subjects, respectively, using just nine EEG channels. The paper also provides insights into the subspace model's scalability to unseen tasks and individuals during training and the number of channels needed for subspace modeling.
\end{abstract}


%
\IEEEpeerreviewmaketitle
\section{Introduction}
\label{sec:intro}
Person recognition using EEG is an emerging technology. Previous studies in EEG-based person recognition have been constrained to a particular task or condition.  Several elicitation protocols have been proposed for EEG-based person recognition.  A detailed review of these different protocols and their performance can be found in \cite{Campisi2014, DelPozoBanos2014, Gui2019}.  

Multiple factors suggest that the EEG can contain signatures \cite{Berkhout1968, VANDIS197987} that help to uniquely identify individuals irrespective of the task, condition, or state of the brain. These include genetic differences between individuals, compounded by neural plasticity due to environmental factors and learning \cite{sur2005patterning}, which help specify neuronal connections and brain activity that are reflected in the EEG. The focus of the present work is to identify individuals independent of the task or condition across recording sessions. To this end, the systems proposed in this paper build upon and extend existing state-of-the-art text-independent speaker recognition techniques, namely, the $i$-vector system \cite{dehak2011front} and the  $x$-vector system \cite{Snyder2018XVectorsRD}. Preliminary results on the proposed approaches were presented in  \cite{kumar2019subspace}, and are encouraging. This paper provides a consolidated analysis of these systems and shows that the proposed modifications are better than simple early and late fusion techniques used for modeling channel information.

For the work presented in this paper, a database was systematically collected with different elicitation protocols. A $128$-channel EEG system was used for this purpose.  EEG data were obtained from $30$ healthy volunteers while they performed various tasks. In addition, we also use a large clinical dataset with $920$ subjects, a subset of an openly available EEG dataset collected in a clinical environment \cite{Obeid2016}. This clinical dataset was recorded with clinical tasks which do not include any standard data elicitation protocols. Using these two diverse datasets with multiple tasks and sessions, we show evidence for task-independent person-specific signatures in EEG.

EEG analysis requires adequate spatial sampling to capture the functionality of the brain. Our work suggests that person-specific information is observed in signals from all regions of the brain suggesting that high spatial-resolution may not be essential. Although previous works have explored different sets of channels for task-dependent EEG biometrics, a systematic study of the spatial resolution needed  for task-independent EEG biometrics is still lacking. Using the $128$ channels EEG dataset, this paper systematically compares various models built with different subsets of sensor placement. Different spatial subsampling methods are examined to find the best set of channels necessary for person recognition.

The organization of the paper is as follows. Prior art on EEG biometrics and our contributions are summarized in the remainder of this section. The details of the baseline and proposed EEG person recognition systems are given in Section~\ref{sec:Methods}. Section~\ref{sec:Datasets} discusses the different datasets used in this paper. The general experimental setup and the features used are outlined in Section~\ref{sec:Experimental_Setup}. The experiments and results are presented in Section~\ref{sec:Results}, followed by a discussion in Section~\ref{sec:Discussion}. Section~\ref{sec:Conclusion} concludes the paper.

\subsection{Related Work}
 Multiple factors such as (i) sessions, (ii) tasks used to elicit subject-specific signatures, (iii) number of channels used and their location, and (iv) the choice of features and classifier have all been shown to influence the performance of EEG biometrics system. This section presents a review of related work addressing the factors mentioned above.

 \subsubsection{Testing across sessions} 
 \label{sec:Testing-across-sessions}
Most previous studies on EEG biometrics have used datasets that have only a single acquisition session \cite{RUIZBLONDET2017}. The exogenous conditions such as impedance between the electrodes and the scalp, minor displacement in electrode location, power supply artifacts, and other factors that vary from session to session can affect both inter- and intra-subject variability of the EEG recordings \cite{Maiorana2018}. Consequently, many recent studies have shown that the performance of EEG biometrics systems is significantly affected by cross-session testing. In \cite{Holler2019}, the EER for $60$ subjects was observed to degrade from $5.34\%$ to $22.0\%$. In \cite{D2018}, without cross-session testing, $100\%$  accuracy was obtained with $40$ subjects, whereas while testing $15$ subjects across sessions, the performance dropped to $86.8\%$.  Without cross-session testing, the performance of the EEG biometric system can be heavily influenced by session-specific conditions. Owing to this, all results in this paper are reported only by evaluating data from sessions unseen during training. 

Table~\ref{tab:1:Literature_Survey} gives a summary of previous works in the literature that have tested EEG-based person recognition across multiple acquisition  sessions. The results in  Table~\ref{tab:1:Literature_Survey} show immense potential for using EEG as a possible biometric. In \cite{maiorana2020deep}, even with an intersession interval of $16$ months, the biometric system is shown to work with an average accuracy of about $91.4\%$.

\begin{table*}[h!t]
	\caption{Literature on EEG biometrics with cross-session evaluation}
	\label{tab:1:Literature_Survey}
	\renewcommand{\arraystretch}{1.3}
	\resizebox{\textwidth}{!}{%
		\begin{tabular}{cclccllllll}
			\hline
			\multicolumn{1}{c|}{Ref.} & \multicolumn{1}{c|}{\begin{tabular}[c]{@{}c@{}}Sessions\\ per-person\end{tabular}} & \multicolumn{1}{c|}{\begin{tabular}[c]{@{}c@{}}Inter-Session \\ Interval\end{tabular}} & \multicolumn{1}{c|}{\begin{tabular}[c]{@{}c@{}}No. of\\ Individuals\end{tabular}} & \multicolumn{1}{l|}{\begin{tabular}[c]{@{}l@{}}No. of \\ Channels\end{tabular}} & \multicolumn{1}{l|}{\begin{tabular}[c]{@{}l@{}}EPs/\\ Tasks\end{tabular}} & \multicolumn{1}{l|}{Feature} & \multicolumn{1}{l|}{\begin{tabular}[c]{@{}l@{}}Duration of\\ EEG segment\end{tabular}} & \multicolumn{1}{l|}{Classifier} & \multicolumn{1}{l|}{\begin{tabular}[c]{@{}l@{}}Channel\\ Handling\end{tabular}} & \multicolumn{1}{l}{Performance} \\ \hhline{*{11}{=}}
			\multicolumn{1}{c|}{\cite{Marcel2007}} & \multicolumn{1}{c|}{3} & \multicolumn{1}{l|}{3 days} & \multicolumn{1}{c|}{9} & \multicolumn{1}{c|}{8} & \multicolumn{1}{l|}{MT} & \multicolumn{1}{l|}{PSD} & \multicolumn{1}{l|}{N.A} & \multicolumn{1}{l|}{UBM-GMM} & \multicolumn{1}{l|}{FC} & \multicolumn{1}{l}{HTER = 36.2\%} \\ \hline
			\multicolumn{1}{c|}{\cite{napflin2007test}} & \multicolumn{1}{c|}{2} & \multicolumn{1}{l|}{12 - 15 months} & \multicolumn{1}{c|}{20} & \multicolumn{1}{c|}{3} & \multicolumn{1}{l|}{REC} & \multicolumn{1}{l|}{PSD based} & \multicolumn{1}{l|}{290s (median)} & \multicolumn{1}{l|}{LR} & \multicolumn{1}{l|}{FC} & \multicolumn{1}{l}{ACC = 88\%} \\ \hline
			\multicolumn{1}{c|}{\cite{napflin2008test}} & \multicolumn{1}{c|}{2} & \multicolumn{1}{l|}{12 - 15 months} & \multicolumn{1}{c|}{20} & \multicolumn{1}{c|}{3} & \multicolumn{1}{l|}{WM} & \multicolumn{1}{l|}{PSD based} & \multicolumn{1}{l|}{189s (median)} & \multicolumn{1}{l|}{LR} & \multicolumn{1}{l|}{FC} & \multicolumn{1}{l}{ACC = 88\%} \\ \hline
			\multicolumn{1}{c|}{\cite{Brigham2010}} & \multicolumn{1}{c|}{4} & \multicolumn{1}{l|}{N.A} & \multicolumn{1}{c|}{6} & \multicolumn{1}{c|}{106} & \multicolumn{1}{l|}{IS} & \multicolumn{1}{l|}{AR} & \multicolumn{1}{l|}{N.A} & \multicolumn{1}{l|}{SVM} & \multicolumn{1}{l|}{FC} & \multicolumn{1}{l}{ACC = 78.6\% to 99.8\%} \\ \hline
			\multicolumn{1}{c|}{\cite{kostilek2012eeg}} & \multicolumn{1}{c|}{2} & \multicolumn{1}{l|}{1 year} & \multicolumn{1}{c|}{9} & \multicolumn{1}{c|}{53} & \multicolumn{1}{l|}{MM} & \multicolumn{1}{l|}{AR} & \multicolumn{1}{l|}{60s} & \multicolumn{1}{l|}{MD} & \multicolumn{1}{l|}{FC} & \multicolumn{1}{l}{ACC = 64.7\% to 77.8\%} \\ \hline
			\multicolumn{1}{c|}{\cite{Wang2016}} & \multicolumn{1}{c|}{2} & \multicolumn{1}{l|}{1 week (min)} & \multicolumn{1}{c|}{4} & \multicolumn{1}{c|}{128} & \multicolumn{1}{l|}{REC} & \multicolumn{1}{l|}{CWT} & \multicolumn{1}{l|}{2s} & \multicolumn{1}{l|}{k-NN} & \multicolumn{1}{l|}{FC} & \multicolumn{1}{l}{ACC = 92.58\%} \\ \hline
			\multicolumn{1}{c|}{\multirow{2}{*}{\cite{ARMSTRONG201559}}} & \multicolumn{1}{c|}{2} & \multicolumn{1}{l|}{5 - 40 days} & \multicolumn{1}{c|}{15} & \multicolumn{1}{c|}{\multirow{2}{*}{1}} & \multicolumn{1}{l|}{\multirow{2}{*}{\begin{tabular}[c]{@{}l@{}}N400\\ {\cite{kutas1980reading}}\end{tabular}}} & \multicolumn{1}{l|}{\multirow{2}{*}{ERP}} & \multicolumn{1}{l|}{\multirow{2}{*}{1.1s x 100}} & \multicolumn{1}{l|}{\multirow{2}{*}{Cross Corr.}} & \multicolumn{1}{l|}{N.A} & \multicolumn{1}{l}{ACC = 89\%} \\ \cline{2-4} \cline{10-11} 
			\multicolumn{1}{c|}{} & \multicolumn{1}{c|}{3} & \multicolumn{1}{l|}{4 - 6 months} & \multicolumn{1}{c|}{9} & \multicolumn{1}{c|}{} & \multicolumn{1}{l|}{} & \multicolumn{1}{l|}{} & \multicolumn{1}{l|}{} & \multicolumn{1}{l|}{} & \multicolumn{1}{l|}{N.A} & \multicolumn{1}{l}{ACC = 93.0\%} \\ \hline
			\multicolumn{1}{c|}{\cite{RUIZBLONDET2017}} & \multicolumn{1}{c|}{2} & \multicolumn{1}{l|}{9 months (avg)} & \multicolumn{1}{c|}{20} & \multicolumn{1}{c|}{26} & \multicolumn{1}{l|}{VEP} & \multicolumn{1}{l|}{ERP} & \multicolumn{1}{l|}{N.A} & \multicolumn{1}{l|}{Cross Corr.} & \multicolumn{1}{l|}{Voting} & \multicolumn{1}{l}{ACC = 100\%} \\ \hline
			\multicolumn{1}{c|}{\cite{Das2015}} & \multicolumn{1}{c|}{3} & \multicolumn{1}{l|}{25 - 49 days} & \multicolumn{1}{c|}{50} & \multicolumn{1}{c|}{17} & \multicolumn{1}{l|}{VEP} & \multicolumn{1}{l|}{ERP} & \multicolumn{1}{l|}{600ms x 50} & \multicolumn{1}{l|}{CS} & \multicolumn{1}{l|}{SF} & \multicolumn{1}{l}{EER(V) = 10\% to 15\%} \\ \hline
			\multicolumn{1}{c|}{\cite{Holler2019}} & \multicolumn{1}{c|}{2} & \multicolumn{1}{l|}{2 weeks} & \multicolumn{1}{c|}{60} & \multicolumn{1}{c|}{27} & \multicolumn{1}{l|}{REC} & \multicolumn{1}{l|}{Multiple} & \multicolumn{1}{l|}{20.5s} & \multicolumn{1}{l|}{Cross Corr.} & \multicolumn{1}{l|}{FC} & \multicolumn{1}{l}{EER(V) = 22\%} \\ \hline
			
			\multicolumn{1}{c|}{\cite{yu2019eeg}} & \multicolumn{1}{c|}{2} & \multicolumn{1}{l|}{N.A} & \multicolumn{1}{c|}{8} & \multicolumn{1}{c|}{9} & \multicolumn{1}{l|}{SSVEP} & \multicolumn{1}{l|}{PSD based} & \multicolumn{1}{l|}{1s} & \multicolumn{1}{l|}{CNN} & \multicolumn{1}{l|}{FC} & \multicolumn{1}{l}{ACC $\approx$ 97\%} \\ \hline
			
			\multicolumn{11}{c}{Literature on Multi-Session-Multi-Task EEG Biometrics} \\ \hline
			\multicolumn{1}{l|}{\multirow{2}{*}{\cite{la2013repeatability}}} & \multicolumn{1}{c|}{\multirow{2}{*}{2}} & \multicolumn{1}{l|}{\multirow{2}{*}{1 - 3 weeks}} & \multicolumn{1}{c|}{\multirow{2}{*}{9}} & \multicolumn{1}{c|}{\multirow{2}{*}{3}} & \multicolumn{1}{l|}{REC} & \multicolumn{1}{l|}{\multirow{2}{*}{AR}} & \multicolumn{1}{l|}{\multirow{2}{*}{60s}} & \multicolumn{1}{l|}{\multirow{2}{*}{\begin{tabular}[c]{@{}l@{}}Linear\\ Classifier\end{tabular}}} & \multicolumn{1}{l|}{\multirow{2}{*}{FC}} & \multicolumn{1}{l}{ACC = 100\%} \\ \cline{6-6} \cline{11-11} 
			\multicolumn{1}{l|}{} & \multicolumn{1}{c|}{} & \multicolumn{1}{l|}{} & \multicolumn{1}{c|}{} & \multicolumn{1}{c|}{} & \multicolumn{1}{l|}{REO} & \multicolumn{1}{l|}{} & \multicolumn{1}{l|}{} & \multicolumn{1}{l|}{} & \multicolumn{1}{l|}{} & \multicolumn{1}{l}{ACC = 90.3\%} \\ \hline
			\multicolumn{1}{l|}{\multirow{2}{*}{\cite{Maiorana2016a}}} & \multicolumn{1}{c|}{\multirow{2}{*}{2}} & \multicolumn{1}{l|}{\multirow{2}{*}{1 month}} & \multicolumn{1}{c|}{\multirow{2}{*}{30}} & \multicolumn{1}{c|}{19} & \multicolumn{1}{l|}{REC} & \multicolumn{1}{l|}{\multirow{2}{*}{\begin{tabular}[c]{@{}l@{}}PCA\\ based\end{tabular}}} & \multicolumn{1}{l|}{\multirow{2}{*}{5s}} & \multicolumn{1}{l|}{\multirow{2}{*}{\begin{tabular}[c]{@{}l@{}}L1, L2 \&\\ CS\end{tabular}}} & \multicolumn{1}{l|}{\multirow{2}{*}{FC}} & \multicolumn{1}{l}{ACC = 87.9\%} \\ \cline{5-6} \cline{11-11} 
			\multicolumn{1}{l|}{} & \multicolumn{1}{c|}{} & \multicolumn{1}{l|}{} & \multicolumn{1}{c|}{} & \multicolumn{1}{c|}{} & \multicolumn{1}{l|}{REO} & \multicolumn{1}{l|}{} & \multicolumn{1}{l|}{} & \multicolumn{1}{l|}{} & \multicolumn{1}{l|}{} & \multicolumn{1}{l}{ACC = 75.4\%} \\ \hline
			\multicolumn{1}{l|}{\multirow{2}{*}{\cite{Maiorana2016b}}} & \multicolumn{1}{c|}{\multirow{2}{*}{3}} & \multicolumn{1}{l|}{\multirow{2}{*}{1 month}} & \multicolumn{1}{c|}{\multirow{2}{*}{50}} & \multicolumn{1}{c|}{\multirow{2}{*}{19}} & \multicolumn{1}{l|}{REC} & \multicolumn{1}{l|}{\multirow{2}{*}{\begin{tabular}[c]{@{}l@{}}AR,\\ PSD\end{tabular}}} & \multicolumn{1}{l|}{\multirow{2}{*}{90s}} & \multicolumn{1}{l|}{\multirow{2}{*}{\begin{tabular}[c]{@{}l@{}}L1, L2 \&\\ CS\end{tabular}}} & \multicolumn{1}{l|}{\multirow{2}{*}{SF}} & \multicolumn{1}{l}{ACC = 90.8\%} \\ \cline{6-6} \cline{11-11} 
			\multicolumn{1}{l|}{} & \multicolumn{1}{c|}{} & \multicolumn{1}{l|}{} & \multicolumn{1}{c|}{} & \multicolumn{1}{c|}{} & \multicolumn{1}{l|}{REO} & \multicolumn{1}{l|}{} & \multicolumn{1}{l|}{} & \multicolumn{1}{l|}{} & \multicolumn{1}{l|}{} & \multicolumn{1}{l}{ACC = 85.6\%} \\ \hline
			\multicolumn{1}{l|}{\multirow{2}{*}{\cite{delpozo2018evidence}}} & \multicolumn{1}{c|}{2} & \multicolumn{1}{l|}{N.A} & \multicolumn{1}{c|}{10} & \multicolumn{1}{c|}{18} & \multicolumn{1}{l|}{VEP} & \multicolumn{1}{l|}{\multirow{2}{*}{PSD}} & \multicolumn{1}{l|}{1s} & \multicolumn{1}{l|}{\multirow{2}{*}{NB}} & \multicolumn{1}{l|}{\multirow{2}{*}{FC}} & \multicolumn{1}{l}{ACC = 41.7\% to 42.9\%} \\ \cline{2-6} \cline{8-8} \cline{11-11} 
			\multicolumn{1}{l|}{} & \multicolumn{1}{c|}{2} & \multicolumn{1}{l|}{N.A} & \multicolumn{1}{c|}{5} & \multicolumn{1}{c|}{6} & \multicolumn{1}{l|}{MT} & \multicolumn{1}{l|}{} & \multicolumn{1}{l|}{2s} & \multicolumn{1}{l|}{} & \multicolumn{1}{l|}{} & \multicolumn{1}{l}{ACC = 73.5\% to 82.1\%} \\ \hline
			
			\multicolumn{1}{l|}{\multirow{4}{*}{\cite{Maiorana2018}}} & \multicolumn{1}{c|}{\multirow{4}{*}{5}} & \multicolumn{1}{l|}{\multirow{4}{*}{}} & \multicolumn{1}{c|}{\multirow{4}{*}{45}} & \multicolumn{1}{c|}{\multirow{4}{*}{19}} & \multicolumn{1}{l|}{REC} & \multicolumn{1}{l|}{\multirow{4}{*}{AR}} & \multicolumn{1}{l|}{5s} & \multicolumn{1}{l|}{\multirow{4}{*}{HMM}} & \multicolumn{1}{l|}{\multirow{4}{*}{Voting}} & \multicolumn{1}{l}{EER(V) = 6.6\%} \\ \cline{6-6} \cline{8-8} \cline{11-11} 
			\multicolumn{1}{l|}{} & \multicolumn{1}{c|}{} & \multicolumn{1}{l|}{Min $\approx$ 1 week} & \multicolumn{1}{c|}{} & \multicolumn{1}{c|}{} & \multicolumn{1}{l|}{REO} & \multicolumn{1}{l|}{} & \multicolumn{1}{l|}{5s} & \multicolumn{1}{l|}{} & \multicolumn{1}{l|}{} & \multicolumn{1}{l}{EER(V) = 10.6\%} \\ \cline{6-6} \cline{8-8} \cline{11-11} 
			\multicolumn{1}{l|}{} & \multicolumn{1}{c|}{} & \multicolumn{1}{l|}{Max $\approx$ 36 months} & \multicolumn{1}{c|}{} & \multicolumn{1}{c|}{} & \multicolumn{1}{l|}{MT} & \multicolumn{1}{l|}{} & \multicolumn{1}{l|}{5s} & \multicolumn{1}{l|}{} & \multicolumn{1}{l|}{} & \multicolumn{1}{l}{EER(V) = 10.7\%} \\ \cline{6-6} \cline{8-8} \cline{11-11} 
			\multicolumn{1}{l|}{} & \multicolumn{1}{c|}{} & \multicolumn{1}{l|}{} & \multicolumn{1}{c|}{} & \multicolumn{1}{c|}{} & \multicolumn{1}{l|}{IS} & \multicolumn{1}{l|}{} & \multicolumn{1}{l|}{5s} & \multicolumn{1}{l|}{} & \multicolumn{1}{l|}{} & \multicolumn{1}{l}{EER(V) = 9.7\%} \\ \hline
			
			\multicolumn{1}{l|}{\multirow{6}{*}{\cite{maiorana2020deep}}} & \multicolumn{1}{c|}{\multirow{6}{*}{5}} & \multicolumn{1}{l|}{\multirow{6}{*}{}} & \multicolumn{1}{c|}{\multirow{6}{*}{45}} & \multicolumn{1}{c|}{\multirow{6}{*}{19}} & \multicolumn{1}{l|}{REC} & \multicolumn{1}{l|}{\multirow{6}{*}{}} & \multicolumn{1}{l|}{5s} & \multicolumn{2}{c|}{\multirow{6}{*}{CNNs \& RNNs}}  & \multicolumn{1}{l}{ACC = 91.4\%} \\ \cline{6-6} \cline{8-8} \cline{11-11} 
			\multicolumn{1}{l|}{} & \multicolumn{1}{c|}{} & \multicolumn{1}{l|}{} & \multicolumn{1}{c|}{} & \multicolumn{1}{c|}{} & \multicolumn{1}{l|}{REO} & \multicolumn{1}{l|}{} & \multicolumn{1}{l|}{5s} & \multicolumn{2}{l|}{} & \multicolumn{1}{l}{ACC = 81.9\%} \\ 
			\cline{6-6} \cline{8-8} \cline{11-11} 
			\multicolumn{1}{l|}{} & \multicolumn{1}{c|}{} & \multicolumn{1}{l|}{Min $\approx$ 1 week} & \multicolumn{1}{c|}{} & \multicolumn{1}{c|}{} & \multicolumn{1}{l|}{MI} & \multicolumn{1}{l|}{AR~+} & \multicolumn{1}{l|}{5s} & \multicolumn{2}{l|}{} & \multicolumn{1}{l}{ACC = 86.2\%} \\ 
			\cline{6-6} \cline{8-8} \cline{11-11} 
			\multicolumn{1}{l|}{} & \multicolumn{1}{c|}{} & \multicolumn{1}{l|}{Max $\approx$ 16 months} & \multicolumn{1}{c|}{} & \multicolumn{1}{c|}{} & \multicolumn{1}{l|}{MT} & \multicolumn{1}{l|}{MFCC} & \multicolumn{1}{l|}{5s} & \multicolumn{2}{l|}{} & \multicolumn{1}{l}{ACC = 84.1\%} \\ 
			\cline{6-6} \cline{8-8} \cline{11-11} 
			\multicolumn{1}{l|}{} & \multicolumn{1}{c|}{} & \multicolumn{1}{l|}{} & \multicolumn{1}{c|}{} & \multicolumn{1}{c|}{} & \multicolumn{1}{l|}{VS} & \multicolumn{1}{l|}{} & \multicolumn{1}{l|}{5s} & \multicolumn{2}{l|}{} & \multicolumn{1}{l}{ACC = 81.6\%} \\ 
			\cline{6-6} \cline{8-8} \cline{11-11} 
			\multicolumn{1}{l|}{} & \multicolumn{1}{c|}{} & \multicolumn{1}{l|}{} & \multicolumn{1}{c|}{} & \multicolumn{1}{c|}{} & \multicolumn{1}{l|}{IS} & \multicolumn{1}{l|}{} & \multicolumn{1}{l|}{5s} & \multicolumn{2}{l|}{} & \multicolumn{1}{l}{ACC = 86.1 \%} \\ \hline
			
			\multicolumn{11}{l}{\begin{tabular}[c]{@{}l@{}}Abbreviations: EP - Elicitation Protocol, MI - Motor Imagery,  REC - Resting Eye Closed, WM - Working Memory, IS - Imagined Speech, MM - Motor Movement, \\  VEP - Visually  Evoked Potential, REC - Resting Eye Open, MT - Mental Tasks, SSVEP -  Steady State Visually Evoked Potential, AR - Auto-Regressive,  \\ PSD - Power Spectral Density, CWT - Continuous Wavelet Transform, ERP - Event-Related Potential, LR - Linear Regression, GMM - Gaussian Mixture Model, \\ SVM - Support Vector Machine, MD - Mahalanobis Distance, CS - Cosine Similarity, NB -Gaussian Naive Bayes Classifier, CNN - Convolutional Neural Network,\\  HMM - Hidden Markov Model, FC - Feature Concatenation, SF - Score Fusion, HTER - Half Total Error Rate, ACC - Accuracy\\ EER(V) - Equal Error Rate computed in an verification framework. \end{tabular}}
		\end{tabular}%
	}
\end{table*}

\subsubsection{Tasks used for EEG biometrics}
\label{sec:Tasks-used-for-EEG-biometrics}
From Table~\ref{tab:1:Literature_Survey}, it is important to observe that different studies have employed different elicitation protocols. The primary interest involved in studying different elicitation protocols is that individuals can have a  distinctive signature for a given task, which can be leveraged to identify them. However, in Table~\ref{tab:1:Literature_Survey}, almost every elicitation protocol has demonstrated success in EEG person recognition across sessions. This observation suggests that person-specific signatures are present under all task conditions, and hence biometric systems need not be designed for a particular elicitation protocol. Consequently, EEG biometric systems have been shown to work on multiple tasks or conditions by training and testing on each task separately \cite{la2013repeatability, Maiorana2016a, Maiorana2016b, Maiorana2018, maiorana2020deep}. Since these systems are trained only on a specific task, they may not scale to other protocols.  Building a task-independent EEG biometric system can eliminate the constraint of using an elicitation protocol. Recent studies have explored the  task-independent nature of EEG biometrics using single-session data and a small set of tasks \cite{delpozo2018evidence,tran2019eeg, wang2019convolutional,fraschini2019robustness,kong2018task}. This lack of cross-session testing is stated as a significant limitation \cite{delpozo2018evidence,wang2019convolutional}. Cross session testing is important because the session-specific factors are known to influence the EEG biometrics (Section~\ref{sec:Testing-across-sessions}). In \cite{delpozo2018evidence}, the results of using different tasks for training and testing across two sessions with $<=10$ subjects on two datasets are presented. The first dataset has a resting state and four mental subtasks data. The second dataset has two tasks, which are EEG captured with self and non-self images as stimuli. These subtasks are not known to influence the EEG significantly. This has been stated as a major limitation in \cite{delpozo2018evidence}. In \cite{D2018}, we studied task-independent nature on $15$ subjects using five different elicitation protocols but limited to the closed eye condition.  The present paper provides a detailed  analysis of the task-independent nature of biometric signatures in EEG by using a dataset collected using $12$ different elicitation protocols with both auditory and visual stimuli. 

\subsubsection{Channels used for EEG biometric}
\label{sec:Channels-used-for-EEG-biometric}
In prior work, multiple techniques have been used to reduce the number of channels needed for biometric recognition using EEG. This subsampling is essential because increasing the number of channels increases the computational complexity of the biometric system. Some works sample the channels according to the task, for instance, centro-parietal channels for a mental task \cite{Marcel2007} and parietal-occipital lobe for a visual task \cite{yu2019eeg}. In \cite{Wang2016}, principal component analysis (PCA) was used to reduce the number of channels. However, the most widely used technique is to select a subset of channels  based on performance of the biometric system \cite{ARMSTRONG201559, RUIZBLONDET2017, Das2015,Maiorana2016a,la2013repeatability,Maiorana2016b,Maiorana2018}.

In \cite{Maiorana2016a,Maiorana2016b}, it is shown that the performance  of the electrodes from the occipital lobe is better for the eyes-closed recording owing to the alpha activity in the visual cortex. However, for the recordings with the eyes-open condition, the performance was similar to the frontal, central lobe. In \cite{napflin2007test,napflin2008test}, three electrodes were chosen such that they are spatially located far from each other. The location of the electrode also accounts for the spatial variation. The dominant frequencies at the frontal lobe are generally lower in other lobes such as parietal, occipital \cite{henry2006electroencephalography}. In this work, we initially sample $9$ channels from the standard $10$-$20$ EEG system such that they are spatially apart and cover different lobes of the brain. This selection is justified by empirically studying different configurations of channels in Section~\ref{sec:channel_effect}.

\subsubsection{Features and Classifiers used for EEG Biometrics}
\label{sec:Features-and-Classifiers-used-for-EEG-Biometrics}
The most commonly used features include spectral analysis using discrete Fourier transform (DFT) \cite{Marcel2007,napflin2007test,napflin2008test, Maiorana2016a, Maiorana2016b, delpozo2018evidence} or continuous wavelet transform (CWT) \cite{Wang2016} and 
autoregressive (AR) coefficients \cite{ la2013repeatability,  Brigham2010, Maiorana2016b, Maiorana2018}. Besides, few studies have averaged the EEG signal across multiple trials and have used the event-related-potential (ERP) as features \cite{RUIZBLONDET2017,ARMSTRONG201559,Das2015}. Using ERP is not feasible in task-independent EEG biometrics. In the case of AR features, a small change in the estimated coefficients can change the location of the roots in the {\it z}-domain. This can affect the frequency spectrum of the EEG signal quite significantly. The raw power spectral density (PSD) estimated on short windows has been shown to identify subjects across sessions in \cite{D2018}. Hence, in this paper, raw PSD estimated over short windows are used as features for recognizing
 individuals.

Table~\ref{tab:1:Literature_Survey} shows that longer the duration of EEG signal used, better the performance of EEG biometrics. \cite{la2013repeatability}  achieved $100\%$ accuracy on $9$ individuals using $60s$ of EEG data. \cite{Das2015} used ERP averaged across multiple trials to achieve $100\%$ recognition on $20$ participants. The best performance obtained for short duration of EEG, such as $5s$ is accuracy of $91.4\%$ for $45$ subjects in \cite{maiorana2020deep}. A short duration of $15$s and a long duration of $60$s are both used to evaluate systems in this paper.

Most of the prior-work discussed in Table~\ref{tab:1:Literature_Survey} have used relatively simple classification/verification methods like SVM \cite{Brigham2010}, Bayes classification \cite{delpozo2018evidence}, scoring techniques like L1 distance, cosine similarity, Mahalanobis distance \cite{ARMSTRONG201559,Das2015,Holler2019,Maiorana2016a,Maiorana2016b, RUIZBLONDET2017} or a nearest neighbor classifier \cite{Wang2016}. However, the challenge involved in task-independent  EEG person recognition is minimizing the information about the task or state of the brain and the session related information present in EEG. This problem is similar to text-independent speaker recognition. In speaker recognition, the primary assumption is that the speaker information is present in the entire signal in addition to that of phoneme and channel information. Consequently, subspace techniques such as $i$-vector \cite{dehak2011front} and $x$-vector \cite{Snyder2018XVectorsRD} were proposed for speaker recognition. These models try to encode the speaker information present in the speech signal on to a compact vector representation. The derivation of $i$-vector representation  is based on an expectation-maximization based algorithm introduced in \cite{dehak2011front}  using distributional statistics of a data model (Gaussian Mixture Model). $x$-vector \cite{Snyder2018XVectorsRD} is a more  recent deep neural network (DNN) based representation  that has outperformed $i$-vectors in speaker recognition tasks. Both methods assume speaker information is present in the entirety of the speech signal and estimate various statistics across time and project them on to a lower-dimensional space. This paper proposes modifications to both the $i$-vector system and the $x$-vector system to take advantage of parallel information available across multiple EEG channels.

\subsection{Contributions} 
\label{sec:contrib}

In \cite{kumar2019subspace}, modified versions of the $i$-vector and the $x$-vector systems to recognize individuals using multi-channel EEG were proposed by the authors. In \cite{kumar2019subspace}, only preliminary results were discussed with no inter-task or inter-subject analysis. This manuscript presents an in-depth analysis of the subspace systems proposed in \cite{kumar2019subspace} using two datasets. Dataset $1$ is a modified version of the dataset used in \cite{kumar2019subspace} with $30$ subjects and $12$ different tasks. Dataset $1$ is the primary dataset that has been used in all the experiments. Wherever possible, we also show our results on a large publicly available dataset with $920$ subjects.  The following are the primary contributions in this paper.

\begin{itemize}

\item  This paper builds upon subspace systems introduced in \cite{kumar2019subspace} and proposes a novel system that combines the $i$-vector and the $x$-vector representations. The proposed systems are shown to outperform the current state-of-the-art deep neural networks based classifiers on both datasets.


\item This paper uses dataset 1, collected with $12$ significantly different and diverse elicitation protocols, to test the task-independence with mismatched testing conditions. We further categorize the tasks based on harder criteria such as open and closed eye conditions and study the influence on classification when training and test sets are disjoint (eye open vs. eye closed). Testing across sessions on a challenging set of tasks/conditions, this paper builds evidence for task-independent individual-specific signatures in EEG.


\item  In addition to testing the generalizability across tasks, we show that, for both datasets, the subspace approach scales to subjects unseen during training with degradation in performance.

\item Using dataset 1, collected using a $128$-channel EEG system, various channel sub-sampling techniques are explored to achieve better performance with a task-independent setup.
 
\item The baseline versions of the subspace systems discussed in this paper were designed for voice biometrics, where a single channel is typical. In \cite{kumar2019subspace}, these subspace systems were modified to process information from multiple channels by concatenating channel-wise statistics at an intermediate processing level. However, the channel information can also be modeled by either early concatenation of features or late fusion of scores from systems built using individual channels. This paper shows that in the $i$-vector and $x$-vector subspace context, the proposed modification in \cite{kumar2019subspace} is better than the direct early and late fusion techniques to model the data from different models. This result is shown on both datasets.

\end{itemize}

\section{Proposed and Baseline Systems}
\label{sec:Methods}
\subsection{Universal Background Model-Gaussian Mixture Model (UBM-GMM)}
\label{sec:ubm}
The UBM-GMM system proposed in \cite{reynolds2000speaker} is a precursor to the $i$-vector system. A Gaussian mixture model (GMM)  is trained using data pooled from the training sessions of all the individuals. This GMM is also called a universal background model (UBM) as it is estimated using multiple subjects and acts as a reference to the person-specific models. The UBM is then converted to person-specific models by maximum-a-posteriori (MAP) adaptation on person-specific data. While testing, the score is calculated as the log of likelihood ratio between the adapted person-specific model and UBM. A detailed description of the UBM-GMM system for speaker recognition task can be found in \cite{reynolds2000speaker}.

Both EEG and speech are essentially time series signals. Hence many studies in EEG biometrics literature have explored UBM-GMM based techniques \cite{D2018, Marcel2007,  nguyen2013eeg, Altahat2015, Davis2015}. We use this well-studied system as a baseline system for evaluating subspace systems. In this implementation, for building the UBM-GMM system, features were pooled from all the available channels. Hence, this system does not model the channels explicitly.

\subsection{Modified-$i$-vector}
\label{sec:ivector}
$i$-vector is a powerful speech signal representation that has led to state-of-the-art speaker recognition systems \cite{dehak2011front} and has demonstrated the ability to model person-specific information in a lower-dimensional space. The $i$-vector space is a subspace of the UBM space defined as follows:

\begin{align}
\bar{M} = \bar{m} + \mathbf{T}\bar{w}
\end{align}
where $\bar{M}$ is the supervector representing an EEG segment, $\bar{m}$ is the UBM supervector, $\mathbf{T}$ is the total variability matrix that defines the subspace, and $\bar{w}$ is the lower dimensional $i$-vector. The supervector is a vector of concatenated means from the UBM or adapted models. Hence, the dimension of the supervector is $Kd \times1$, where $K$ is the number of Gaussian mixtures, and $d$ is the dimension of the input power spectral density (PSD) feature vector. The $\mathbf{T}$-matrix is of dimension $Kd \times R$, where $R$ is the  dimension of the subspace. $R$ is an empirically determined hyper-parameter of the $i$-vector system.   

Let,
\begin{equation}
X = \left\lbrace \; \bar{x}_{n}^{c} \; | \; n = 1 \; \text{to} \; N \; \& \;  c = 1\; \text{to} \; C  \right\rbrace 
\end{equation}
denote an EEG segment with $C$ EEG sensors/channels and $N$ feature vectors per channel. A $K$ mixture UBM is trained using EEG segments from multiple subjects. Using the UBM, the zeroth and first order statistics required for estimating the $i$-vector are calculated as given in Equations~\ref{eq:3:zostat} and~\ref{eq:4:fostat}, respectively.

\begin{align}
N_{k}(X)  &= \sum_{c=1}^{C} \sum_{n=1}^{N} \; P(k| \bar{x}_{n}^{c},  \lambda) \label{eq:3:zostat} \\
\bar{F}_{k}(X)  &= \sum_{c=1}^{C} \sum_{n=1}^{N} \; P(k|\bar{x}_{n}^{c},  \lambda) (\bar{x}_{n}^{c} - \bar{m}_k) \label{eq:4:fostat} 
\end{align} 

where $\lambda$ represents UBM parameters, $k$ denotes the mixture ID, and $P(k| \bar{x}_{n}^{c},  \lambda)$ corresponds to the posterior probability of the $k$-th mixture component given the feature vector $\bar{x}_{n}^{c}$. $\bar{m}_k$ is the mean of the $k$-th UBM component. Given the zeroth and first-order statistics, the $i$-vector is estimated as follows
 \begin{align}
 	\bar{w} = \left(\mathbf{I}+\mathbf{T}^{t} \mathbf{\Sigma}^{-\frac{1}{2}}  \mathbf{N}(X) \mathbf{T} \right)^{-1}  \mathbf{T}^{t} \mathbf{\Sigma}^{-1} \bar{F}(X) 
 	\label{eq:5:ivector_dim_reduction}
 \end{align}

where $\bar{F}(X)$ is a supervector obtained by concatenation of $\bar{F}_{k}(X) $ for all $k=1...K$ mixtures. Hence the dimension of supervector $\bar{F}(X)$ is $Kd \times 1$. $\mathbf{\Sigma}$  is a $Kd \times Kd$ block diagonal matrix with $\mathbf{\Sigma}_{k}$ (covariance matrix of $k$-th Gaussian) as blocks along the diagonal. $\mathbf{N}(X)$ is also a block diagonal matrix of dimension $Kd \times Kd$ with $N_k(X)\mathbf{I}$ as diagonal blocks. The expectation-maximization (EM) algorithm for estimating $\mathbf{T}$-matrix in the case of EEG person recognition is the same as that for speech and has been detailed in \cite{dehak2011front, Kenny2005}. This system will be referred to as {\it ``baseline-$i$-vector''} in the rest of the paper. This system has been adopted for EEG biometrics in \cite{Ward2016, Ward2016b}. However, this standard approach is not adequate for multi-channel EEG as it does not explicitly model channel information. To integrate information from different channels in the $i$-vector framework, we proposed a novel way of finding the zeroth and first-order statistics as given in Equation~\ref{eq:6:modzostat} and \ref{eq:7:modfostat}, respectively. 

\begin{align}
N_{kc} (X) &=  \sum_{n=1}^{N} \; P(k\;|\; \bar{x}_{n}^{c}, \; \lambda) \label{eq:6:modzostat}\\
\bar{F}_{kc} (X) &=  \sum_{n=1}^{N} \; P(k\;|\; \bar{x}_{n}^{c}, \; \lambda) (\bar{x}_{n}^{c} - \bar{m}_k) \label{eq:7:modfostat}
\end{align}

In this approach, the UBM is still trained by pooling data from all the channels. However, during statistics estimation, it is done for each channel individually and then concatenated before projecting to the lower dimensional $i$-vector space. Hence the supervector $\bar{F}(X)$ (in Eq~\ref{eq:5:ivector_dim_reduction}) is obtained by concatenating $\bar{F}_{kc}(X) $ for all $k=1...K$ Gaussian mixtures and $c=1...C$ channels. Hence the super vector dimension increases to $KCd \times 1$. Consequently, the dimensions of matrices $\mathbf{\Sigma}$, $\mathbf{N}(X)$, and $\mathbf{T}$ (in Eq~\ref{eq:5:ivector_dim_reduction}) increases to $KCd \times KCd$, $KCd \times KCd$,  and $KCd\times R$, respectively. This system is henceforth referred to as {\it ``modified-$i$-vector''}.  Since the dimension of the supervector is high, for effective estimation of $\mathbf{T}$-matrix, we use a smaller number of mixtures in the UBM compared to the baseline model (Table~\ref{tab:ubm_param}).

After estimating the $i$-vector, a linear transform is applied using linear discriminant analysis (LDA). LDA makes the subspace more discriminative for person-specific signatures and hence improves the performance of the person recognition system. During testing, a cosine similarity classifier discussed in Section~\ref{sec:backend} is used on the LDA projected $i$-vectors.

\subsection{Modified-$x$-vector}
\label{sec:xvector}
$x$-vector is a recent DNN based speech representation approach aimed towards speaker recognition \cite{Snyder2018XVectorsRD}. The $x$-vector system initially operates at the frame level, estimates statistics, and then the final few layers operate at the segment level. This architecture is analogous to the $i$-vector system with UBM acting at the frame level and the $\mathbf{T}$-matrix operating at the segment level. Similar to the ``modified-$i$-vector'' in Section~\ref{sec:ivector}, we remodel the $x$-vector system to handle information from multiple EEG sensors. $x$-vector proposed for speech data uses time-delay neural networks (TDNNs) for modeling temporal context. However, upon experimenting with EEG data, we did not find long term context information to be helpful. Hence, $1$-D convolution is used in place of TDNN for $x$-vector based EEG person recognition systems in this paper.

\begin{figure}[h!t]
	\centering
	\includegraphics[width=0.485\textwidth]{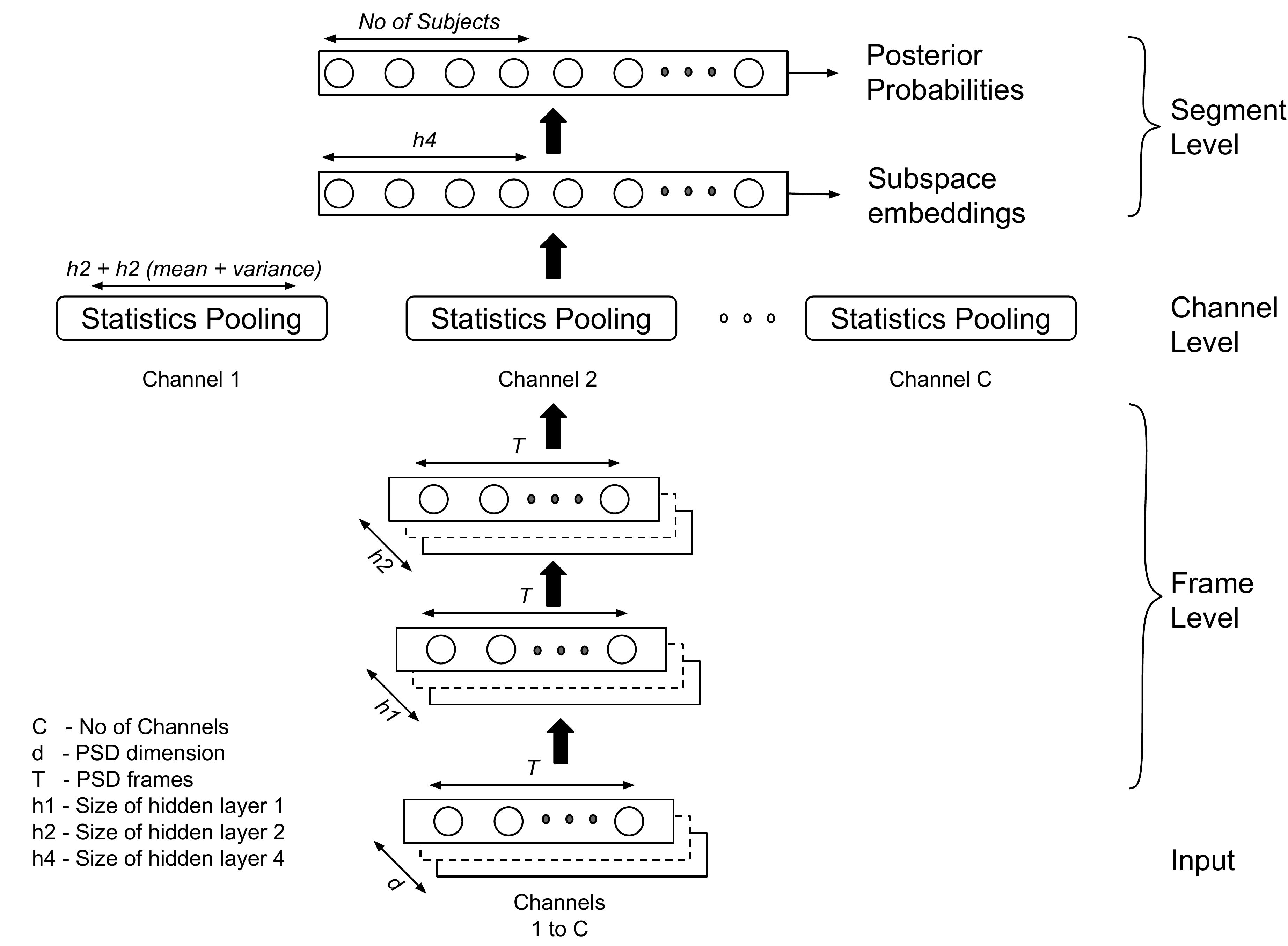}
	\caption{DNN architecture of the modified $x$-vector model for multichannel EEG.}
	\label{fig:arch}
\end{figure}

Figure.~\ref{fig:arch} gives an overview of the $x$-vector architecture modified for multi-channel EEG. Spectrograms from all the channels are provided as input to this model. The model has four hidden layers. The initial two layers are single frame 1-D convolution layers that transform every feature vector of the spectrogram into a higher dimensional space. The third layer is a statistics pooling layer, which estimates the mean and variance for each channel. These statistics are concatenated and reduced to a lower-dimensional representation using the fourth hidden layer. The final output layer is a feed-forward layer with softmax activation. The number of nodes in the output layer is the total number of subjects in the training data. Similar to  \cite{Snyder2018XVectorsRD}, cross-entropy error is used to train the network using Adam optimization \cite{adam}. After training,  the output of the fourth hidden layer is considered as a subspace representation for the EEG segment, also referred to as $x$-vector. This way of estimating $x$-vectors will be henceforth referred to as ``modified-$x$-vector'' system.

The $x$-vector system with a single statistics pooling across all channels in the third hidden layer is identical to $x$-vector proposed for speech and will be used as a baseline. Similar to the ``baseline-$i$-vector'', this system does not take any explicit information about the channels and will be referred to as  ``baseline-$x$-vector'' system.  Testing is performed using a simple cosine similarity classifier after subjecting the $x$-vectors to LDA, as discussed in Section~\ref{sec:backend}.

\subsection{Combined $i$-vector and $x$-vector ($ix$-vector)}
\label{sec:ixvector}
In this section, we propose a novel subspace system that uses both the E-M based modified-$i$-vector (Section~\ref{sec:ivector}) and the DNN based modified-$x$-vector (Section~\ref{sec:xvector}) to recognize individuals from EEG signal. $i$-vectors are subspace embeddings obtained by aligning statistics from a universal background model (UBM) and a large projection (total variability space or T) matrix. Both UBM and the T-matrix are generatively modeled using E-M. On the other hand, $x$-vectors are extracted from a DNN trained discriminatively.  By concatenating embeddings from modified-$i$-vector and modified-$x$-vector, we show that the performance of the individual systems can be significantly improved. This concatenated embeddings system will be henceforth referred to as the  ``$ix$-vector'' system. Section~\ref{sec:backend} details the back-end used to recognize the individuals given the concatenated embeddings.

\subsection{Back-end}
\label{sec:backend}
After estimating the subspace embeddings (Section~\ref{sec:ivector}, ~\ref{sec:xvector},~and~\ref{sec:ixvector}), any classifier can be used for implementing person recognition. Since the focus of this paper is on the subspace technique, we use a simple cosine similarity based recognition system as the back-end. 

Let $\bar{w}_i$ be the subspace vector obtained by projecting all the  available training data of person $i$ and $\bar{w}_{test}$ be the subspace vector under test. The cosine similarity score for $\bar{w}_{test}$ belonging to a person $i$ is calculated as given in Equation~\ref{eq:8:cs}.

\begin{align}
\text{S}_i =  \frac{{\bar{w}_{\text{i}}}^{T} \bar{w}_{\text{test}}}{\norm{\bar{w}_{\text{i}}}\norm{\bar{w}_{\text{test}}}} \label{eq:8:cs}
\end{align} 


\subsection{Other baseline systems}
This section briefly describes the two start-of-the-art DNN based EEG biometrics systems used to compare with our results. The first system is a convolution neural network (CNN) based technique \cite{lawhern2018eegnet}, proposed using multiple CNNs to extract EEG features for BCI applications automatically. This system has also been used for biometric recognition in \cite{maiorana2020deep}. Henceforth, this system will be referred to as the ``EEGNet" system (similar to \cite{lawhern2018eegnet, maiorana2020deep}). The second system is CNN-RNN based system proposed in \cite{maiorana2020deep}, which uses CNNs to convolve across channels and recurrent neural networks (RNNs) or long short-term memory (LSTMs) to convolve across time. This system will be henceforth referred to as the ``CNN-RNN" system.

For the EEGNet system, we have used the implementation provided by the authors (publicly available). The EEGNet system proposed in \cite{lawhern2018eegnet} uses raw EEG signals (time-domain representation) as input.  To enable fair comparison with proposed models, we also show the result for an EEGNet variant that uses a spectrogram or time-frequency representation from each channel as input. For both the EEGNet based systems, filter sizes were fixed at half the sampling rate of giving input (as recommended in \cite{lawhern2018eegnet}), and the number of filters in each layer was fine-tuned. The CNN-RNN system was implemented by the authors similar to the system  $S_{\zeta}$ in \cite{maiorana2020deep}. The filter sizes were kept numerically identical to the model implemented in \cite{maiorana2020deep}, and the number of filters was fine-tuned. 

\section{Datasets}
\label{sec:Datasets}

\subsection{Dataset $1$: $128$-Channel Multi-Task EEG Dataset}
\label{sec:dataset1}
This dataset was collected from $30$ subjects performing multiple tasks. Multiple tasks/elicitation protocols were designed with both open and closed eye conditions to collect this data. Table~\ref{tab:2:experiment_protocols} gives a summary of these tasks and protocols. It is to be noted that all the $30$ subjects did not perform all the $12$ tasks mentioned in Table~\ref{tab:2:experiment_protocols}.  EEG data was collected from each subject, for at least $2$ sessions and at most $5$ sessions. During each session, at most of $4$ tasks from Table~\ref{tab:2:experiment_protocols}  were performed. This dataset is a modified version of the dataset presented in \cite{kumar2019subspace}.

\begin{table*}[h!t]
	\centering
	\setlength{\tabcolsep}{2pt}
	\renewcommand{\arraystretch}{1.3}
	\caption{Data collection protocols for dataset 1.}
	\label{tab:2:experiment_protocols}
	\resizebox{1\textwidth}{!}{
		\begin{tabular}{c|P{30mm}|p{120mm}|c|c}
			
			\hline 
			\multirow{2}{*}{\bf S. No.} & \multirow{1}{*}{\bf Experiment} & \multirow{2}{*}{\bf \hspace{4.3cm} Brief Description of Experiment} & \multirow{1}{*}{\bf No. of} &   \multirow{1}{*}{\bf Total Duration} \\ 
			& \multirow{1}{*}{\bf Name } & & \multirow{1}{*}{\bf Participants }  &  \multirow{1}{*}{\bf  (in Minutes)} \\  \hline
			
			\multicolumn{5}{c}{\bf Experiments conducted with  Closed Eye Condition} \\ \hline
			
			$1$ & Odd Ball Classic & Participants were presented with frequent non-target stimuli and infrequent target stimuli. The target and non-target stimuli consist audio beeps differing in frequency or duration.
			&  $13$ & $5$ hr $2$ mins \\ \hline
			$2$ &Odd Ball Stereo & Similar to S.No $1$. The target and non-target stimuli consist of audio beeps played in left and right ear.
			&  $12$ & $1$ hr $55$ mins \\ \hline
			$3$ & Imagining Binary Answers &  A set of binary questions were presented to participants. They were asked to first imagine the answer and then respond with a mouse click.  & $7$ & $3$ hr $21$ mins \\ 
			\hline
			$4$ & Semantically Opposite Words &  Semantically opposite words such as “yes” and “no” were played to the subject over multiple trials. Subject was instructed to respond with left and right mouse
			clicks depending on the semantics of the word being played  & $4$ & $1$ hr $36$ mins \\ 
			\hline
			$5$ & Familiar and Unfamiliar Words &  The subjects were presented with common words and uncommon words. They were expected to respond with a mouse click on hearing a familiar word. & $6$ & $1$ hr $50$ mins \\ 
			\hline
			$6$ & Proper and Improper Sentences &  Regular and ill-formed sentences were played to subject. The subject was required to respond with mouse click on hearing ill-formed sentences. & $8$ & $1$ hr $52$ mins \\ 
			\hline
			$7$ & Motor and Mental Imaginary & Participants were asked to imagine motor-movements such as left and right fist rotation. For mental imaginary task, they were asked to count numbers in reverse.  & $6$  & $3$ hr $13$ mins
			\\ 
			\hline
			$8$ & Passive Audio & Participants were passively listening to a variety of audio stimuli such as music, sentences, stories, and sounds that trigger attention (for example sound of sirens). & $17$ 
			& $3$ hr $33$ mins
			\\
			\hline
			$9$ & Passive Audio Stereo & Similar to S.No. $7$. The auditory stimuli were always played through only one ear (either left/right) at time using headphones.   & $11$ 
			& $2$ hr $46$ mins
			\\
			\hline
			\multicolumn{5}{c}{\bf Experiments conducted with  Open Eye Condition} \\
			\hline
			$10$ &Odd Ball Visual & Similar to S.No $1$. The target and non-target stimuli consist of visual objects varying in shape and color. & $6$ & $33$ mins \\ \hline
			$11$ & Steady State Visually Evoked Potential & Visual objects flickering at different frequencies were displayed to participants. At the end of each trial, a question about the shape or color of the object was asked.
			&  $12$ & $3$ hr $13$ mins \\ \hline
			$12$ & Passive Audio-Visual & Audio-visual clips were played to the participants. At the end of each clip,  a question was asked based on the stimuli.   
			&  $12$ & $3$ hr $2$ mins\\ \hline
			\multicolumn{5}{l}{\begin{tabular}{p{0.43\textwidth}R{0.57\textwidth}}
					Total number of subjects: $30$ & Total number of subjects with  closed eye recordings: $30$ \\  
					Total duration of the dataset: $31$ hours & Total Number of subjects with  open eye recordings : $14$ \\
					& Total number of subjects with both  open and closed eye recordings on all sessions: $10$\end{tabular}}
	\end{tabular}}
\end{table*}

This dataset was collected by the authors in a laboratory setting using a $128$-channel dense-array EEG system manufactured by Electrical Geodesics, Inc (EGI) \cite{egi}. The Ethics Committee of the Indian Institute of Technology Madras approved this study. All the subjects were informed about the aim and scope of the experiment, and written consent was  obtained to collect the data. The EEG data were recorded at a sampling rate of $250Hz$ with the central electrode $Cz$ as the reference electrode. After collecting the dataset, the artifacts present were removed using \cite{Chang2018}, and bad channels were replaced by spherical spline interpolation \cite{perrin1989spherical} (plugins available with EEG lab toolbox \cite{delorme2004eeglab}). The total duration of this dataset is about $31$ hours, with $3$ sessions per person on average. Further statistics on number of sessions per individual, number of EEG segments and intersession intervals between train and test are given in Table~\ref{tab:2:dataset_stats}. This dataset has been made publicly available at \url{https://www.iitm.ac.in/donlab/cbr/eeg_person_id_dataset/}.

\begin{table*}[h!t]
	\centering
	\setlength{\tabcolsep}{2pt}
	\renewcommand{\arraystretch}{1.1}
	\caption{Statistics of datasets used}
	\label{tab:2:dataset_stats}
	\resizebox{0.7\textwidth}{!}{%
		\begin{tabular}{cccccccccccc}
			\multicolumn{1}{c|}{\multirow{4}{*}{Dataset}} & \multicolumn{4}{c|}{\multirow{1}{*}{Number of sessions }}                                                                                                     & \multicolumn{4}{c|}{\multirow{2}{*}{Time between training and last testing session}}                                                                                          & \multicolumn{3}{c}{\multirow{2}{*}{Number of 15 sec trials}}                                                                                                                                                                                                                                \\ 
			\multicolumn{1}{c|}{}                         & \multicolumn{4}{c|}{per subject}                                                                                                                                                         & \multicolumn{4}{c|}{}                                                                                                                                                         &                                                                \\ \cline{2-12} 
			\multicolumn{1}{c|}{}                         & \multicolumn{1}{c|}{\multirow{2}{*}{Avg}} & \multicolumn{1}{c|}{\multirow{2}{*}{Std}} & \multicolumn{1}{c|}{\multirow{2}{*}{Min}} & \multicolumn{1}{c|}{\multirow{2}{*}{Max}} & \multicolumn{1}{c|}{\multirow{2}{*}{Avg}} & \multicolumn{1}{c|}{\multirow{2}{*}{Std}} & \multicolumn{1}{c|}{\multirow{2}{*}{Min}} & \multicolumn{1}{c|}{\multirow{2}{*}{Max}} & \multicolumn{1}{c|}{\multirow{2}{*}{Train}} & \multicolumn{1}{c|}{\multirow{2}{*}{Validation}} & \multicolumn{1}{c}{\multirow{2}{*}{Test}}                        \\ 
			\multicolumn{1}{c|}{}                         & \multicolumn{1}{c|}{}                     & \multicolumn{1}{c|}{}                     & \multicolumn{1}{c|}{}                     & \multicolumn{1}{c|}{}                     & \multicolumn{1}{c|}{}                     & \multicolumn{1}{c|}{}                     & \multicolumn{1}{c|}{}                     & \multicolumn{1}{c|}{}                     & \multicolumn{1}{c|}{}                       & \multicolumn{1}{c|}{}                            & \multicolumn{1}{c}{}                      
			\\ \hline
			\multicolumn{1}{c|}{1 (30 subjects)}                        & \multicolumn{1}{c|}{3.1}              & \multicolumn{1}{c|}{0.8}              & \multicolumn{1}{c|}{2}               & \multicolumn{1}{c|}{5}             & \multicolumn{1}{c|}{44 Days}              & \multicolumn{1}{c|}{67 Days}              & \multicolumn{1}{c|}{1 Day}                & \multicolumn{1}{c|}{193 Days}             & \multicolumn{1}{c|}{4681}                   & \multicolumn{1}{c|}{559}                         & \multicolumn{1}{c}{2255}                       \\ \hline
			\multicolumn{1}{c|}{2 (920 subjects)}                        & \multicolumn{1}{c|}{3.14}            & \multicolumn{1}{c|}{1.65}            & \multicolumn{1}{c|}{2}               & \multicolumn{1}{c|}{19}            & \multicolumn{1}{c|}{10 Months}            & \multicolumn{1}{c|}{15 Months}            & \multicolumn{1}{c|}{0 Days}               & \multicolumn{1}{c|}{126 Months}            & \multicolumn{1}{c|}{225153}                  & \multicolumn{1}{c|}{42171}                        & \multicolumn{1}{c}{133678}                  \\           
		\end{tabular}%
	}
\end{table*}

\subsection{Dataset 2: Temple University Clinical EEG Dataset}
\label{sec:dataset2}
This dataset is a subset of the Temple University hospital EEG data corpus (TUH-EEG) \cite{Obeid2016}. In this paper we have used the TUH-EEG corpus v.1.1.0 containing over $20,000$ clinical EEG recordings collected from about $14,000$ patients. We preprocess this dataset for evaluating the proposed systems, as described below.

The dataset has data from $7424$ patients with average EEG as reference and $6770$ patients with linked-ears data as reference. Out of this, only $2152$ and $1341$ patients have multi-session data collected using average and linked-ears as reference, respectively. \cite{Obeid2016} presents an analysis of the demographics of the patients. Further for each recording the dataset also has an unstructured clinical report in plain text format. Since abnormal pathological conditions such as seizures can affect subject recognition performance, the recordings that were annotated to have abnormal EEG were first removed. It is to be noted that these annotations for abnormal EEG came along with the original dataset and has been algorithmically generated using \cite{Lopez2017,lopezautomated}. After removing the abnormal recordings, the dataset contained $1033$ and $155$ patients with at least two-sessions with average and linked-ears reference, respectively. For further analysis, only the $1033$ subjects recorded with average reference were chosen owing to the higher number of subjects. To improve the signal-to-noise-ratio, we adopted the methods in \cite{Chang2018} for removal of artifacts and bad-channels (channels with flat or noisy data). In this process, if one of the nine channels considered in our experimental setup (Section~\ref{sec:Experimental_Setup}) turns out to be affected, the corresponding EEG recording was discarded.  After these preprocessing steps, the number of subjects with at least two sessions reduced to $920$ subjects with $2889$ recording sessions. Further detailed statistics on number of sessions per individual, number of EEG segments and intersession intervals between train and test are given in Table~\ref{tab:2:dataset_stats}. 


Clinical tasks such as hyperventilation,  photic simulations, and sleep and wakefulness EEG were used to collect this data. Since these data were collected for clinical purposes, the elicitation protocol has not been standardized across acquisitions.  Also, the dataset does not have any annotations regarding the tasks performed. The set of tasks and the clinical setting makes this dataset distinct from dataset 1.  Given the clinical nature, this dataset is used only to show the scalability of proposed approaches over the baseline on a diverse dataset with a large number of subjects.

\section{General Experimental Setup}
\label{sec:Experimental_Setup}

\subsection{Channels}
\begin{figure}[h!t]
	\centering
	\includegraphics[width=0.284\textwidth]{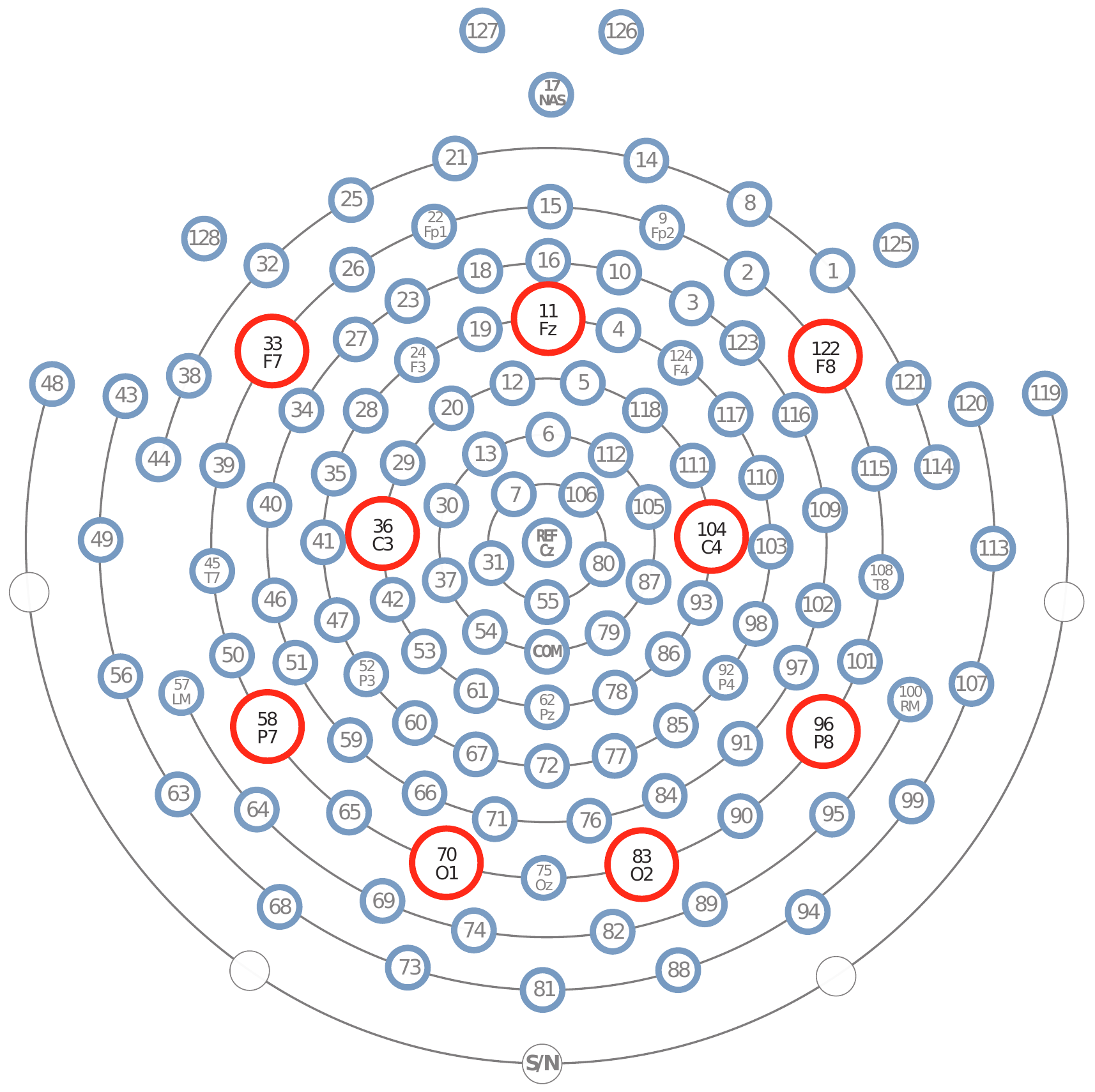}
	\caption{Diagrammatic representation of $128$-channel EEG system. The electrodes with a red outline are the $9$ channels considered for person identification. The sensors in the background denote all the $128$-channel used to collect dataset $1$.}
	\label{fig:2:128channel}
\end{figure}

Dataset $1$ was collected using a $128$-channel EEG system, whereas dataset $2$ was variedly collected using $24$ to $36$ sensors. We initially choose 9 electrodes, namely, Fz, F7, F8, C3, C4, P7, P8, O1, and O2 of the standard $10$-$20$ system to make a common analysis on both datasets.  Later, in Section~\ref{sec:channel_effect}, we analyze the effect of different sets and number of sensors for the proposed models. A diagrammatic representation of these initially chosen $9$ channels with all the $128$ channels as the background is shown in Figure~\ref{fig:2:128channel}. These $9$ electrodes were chosen from the standard $10$-$20$ system such that they cover different regions of the brain, namely, the Frontal, Central, Parietal, and Occipital lobe. This kind of selection covering the entire scalp is essential because different stimuli/tasks elicit different regions of the brain (Section~\ref{sec:channel_effect}). 

\subsection{Features}
\label{sec:ES:Feat}
Power spectral density (PSD) spectrogram is used as the feature. PSD spectrograms are computed in the range of $3$-$30$ Hz  for every channel with a window size of $360$ ms and no overlap. This configuration of PSD features for EEG person recognition was fine-tuned using the UBM-GMM system in \cite{D2018, kumar2019subspace}. 

\subsection{Setup for task-independent person recognition}
In dataset $1$, the EEG signals obtained from various experiments (given in Table.~\ref{tab:2:experiment_protocols}) are divided into segments of $15$s length. This segmentation is done irrespective of the experimental protocol such as whether the person is in the resting state or watching/listening to a stimulus/instruction or doing a specific task. Hence, recognizing individuals from these segments are task independent. We use the same uniform segmentation for dataset $2$,  which was collected for clinical purposes, unlike in a laboratory with formal control of data collection conditions. 

For every individual, the first $60\%$ of the sessions (rounded off to nearest integer) are chosen for training. Of the remaining data, $20\%$  is used for validation, and the rest is used for testing.  For biometric applications, the training data is always collected before the test data. Therefore, the initial few sessions are chosen for training and the remaining for test and validation. Since we divide sessions chronologically, n-fold cross-validation was not performed to analyze the results. Besides, \cite{kumar2019subspace} shows preliminary cross-validation results by randomly dividing sessions for training and testing. It is to be noted that all the results reported in this paper are only on test sessions that are not used during training. Table~\ref{tab:2:dataset_stats} presents the statistics on the time interval between sessions and the number of EEG segments used for training and testing.

\subsection{Evaluation}
The systems are evaluated using two evaluation metrics: rank-1 classification accuracy and equal error rate (EER). In this paper, rank-1 classification accuracy is used to evaluate the proposed systems in a closed-set identification framework, and EER is used for the same in a closed-set verification framework.  For computing classification accuracy, only the maximum score from each EEG segment is used to decide the final class label. For calculating EER, the correct label of a given EEG segment is considered a target, and all the other individuals present in the dataset are considered non-targets. A threshold over all possible scores is used to  determine the EER such that the true positive and false positive rates are the same. Since the subspace models proposed in this paper employ data from all the individuals, the EER computed can only represent a closed-set verification system.  It is to be noted that EER computed in a closed-set verification framework cannot be directly compared to papers (in Table~\ref{tab:1:Literature_Survey}) reporting EER using models that do not use other users’ data, mimicking an open-set verification framework \cite{Das2015,Holler2019,Maiorana2018}. All results in this paper use EER as a metric in addition to rank-1 accuracy because EER is computed using all the score values given the test EEG dataset. Only in case 2 of Table~\ref{tab:7:unseen_subject}, we report EER similar to \cite{Das2015,Holler2019,Maiorana2018}, where we explore the possible extension of the proposed models to an open-set verification framework. 

\section{Experiments and Results}
\label{sec:Results}
\subsection{Performance of proposed systems vs. the baseline systems}
\label{sec:result_1_comp}

  Results of all the baseline and modified systems discussed in Section~\ref{sec:Methods} are compared in Table~\ref{tab:all_result} for both datasets $1$ and $2$. All systems were developed following the common experimental setup given in Section~\ref{sec:Experimental_Setup}. Only the EEGNet system was developed using raw time-domain signals. All the other systems used PSD features (Section~\ref{sec:ES:Feat}). For all the systems, the associated hyper-parameters were fine-tuned using validation data. These fine-tuned layer-wise hidden features used in DNN based systems are given in Table~\ref{tab:dnn_based_config}. Similarly, the number of UBM mixtures and the $i$-vector dimensions are given in Table~\ref{tab:ubm_param} for UBM based models.

\begin{table}[h!t]
	\centering
	\caption{Performance comparison of various system using Accuracy (\%) and EER (\%)}
	\label{tab:all_result}
	\renewcommand{\arraystretch}{1.3}
	\resizebox{0.380\textwidth}{!}{%
		\begin{tabular}{c||c|c|c|c}
			\hline
			\multirow{2}{*}{Systems} & \multicolumn{2}{c|}{\begin{tabular}[c]{@{}c@{}}Dataset 1\\ (30 Subjects)\end{tabular}} & \multicolumn{2}{c}{\begin{tabular}[c]{@{}c@{}}Dataset 2\\ (920 Subjects)\end{tabular}} \\ \cline{2-5} 
			& Acc & EER & Acc & EER \\ \hline \hline
			UBM-GMM & $71.2$ & $10.9$ & $6.02$ & $44.0$ \\ \hline
			baseline-i-vector & $70.5$ & $10.6$ & $9.02$ & $26.8$ \\ \hline
			baseline-x-vector & $67.1$ & $11.3$ & $5.42$ & $29.0$ \\ \hline
			EEGNet \cite{lawhern2018eegnet} & $70.1$  & $9.2$ & $24.0$ & $15.0$ \\ \hline
			EEGNet (PSD) & $77.5$ & $5.79$ & $23.9$ & $15.3$ \\ \hline
			CNN-RNN \cite{maiorana2020deep} & $77.8$ & $6.24$ & $35.0$ & $\mathbf{12.8}$ \\ \hline \hline
			modified-i-vector & $85.1$ & $5.81$ & $30.0$ & $16.5$ \\ \hline
			modified-x-vector & $76.8$ & $8.16$ & $27.2$ & $16.6$ \\ \hline
			ix-vector & $\mathbf{86.4}$ & $\mathbf{5.02}$ & $\mathbf{35.9}$ & $14.2$ \\ \hline
		\end{tabular}%
	}
\end{table}

\begin{table}[h!t]
	\caption{Layer-wise configuration used in DNN based models. (a -  Layer type, b - Number of layer wise feature used for dataset 1 and c - Number of layer wise feature used for dataset 2)}
	\label{tab:dnn_based_config}
	\renewcommand{\arraystretch}{1.3}
	\resizebox{0.48\textwidth}{!}{%
		\begin{tabular}{c|c|ccccccc}
			\hline
			\multicolumn{2}{c|}{System Name} & Layer 1 & Layer 2 & Layer 3 & Layer 4 & Layer 5 & Layer 6 & Output \\ \hline
			\multirow{3}{*}{\begin{tabular}[c]{@{}c@{}}baseline-\\ x-vector\end{tabular}} & (a) & Conv & Conv & FF & - & - & - & FF \\
			& (b) & $1024$ & $1024$ & $160$ & - & - & - & $30$ \\
			& (c) & $1024$ & $128$ & $160$ & - & - & - & $920$ \\ \hline
			\multirow{3}{*}{\begin{tabular}[c]{@{}c@{}}modified-\\ x-vector\end{tabular}} & (a) & Conv & Conv & FF & - & - & - & FF \\
			& (b) & $1024$ & $512$ & $160$ & - & - & - & $30$ \\
			& (c) & $512$ & $128$ & $160$ & - & - & - & $920$ \\ \hline
			\multirow{3}{*}{EEGNet \cite{lawhern2018eegnet}} & (a) & Conv & Conv & Conv & - & - & - & FF \\
			& (b) & $64$ & $128$ & $128$ & - & - & - & $30$ \\
			& (c) & $512$ & $1024$ & $1024$ & - & - & - & $920$ \\ \hline
			\multirow{3}{*}{\begin{tabular}[c]{@{}c@{}}EEGNet \\ (PSD)\end{tabular}} & (a) & Conv & Conv & Conv & - & - & - & FF \\
			& (b) & $64$ & $128$ & $128$ & - & - & - & $30$ \\
			& (c) & $256$ & $512$ & $512$ & - & - & - & $920$ \\ \hline
			\multirow{3}{*}{CNN-RNN \cite{maiorana2020deep}} & (a) & Conv & Conv & Conv & Conv & Conv & Bi-LSTM & FF \\
			& (b) & $32$ & $64$ & $128$ & $256$ & $512$ & $1024$ & $30$ \\
			& (c) & $64$ & $128$ & $256$ & $512$ & $1024$ & $2048$ & $920$ \\ \hline
			\multicolumn{9}{l}{\begin{tabular}[c]{@{}l@{}} Conv - convolution layer,  FF - feed-forward layer,  Bi-LSTM - bidirectional long short term memory.\end{tabular}}
			
		\end{tabular}%
	}
\end{table}

\begin{table}[]
	\caption{ Number of UBM mixtures and the $i$-vector dimensions for UBM based models}
	\label{tab:ubm_param}
	\centering
	\renewcommand{\arraystretch}{1.3}
	\resizebox{0.35\textwidth}{!}{%
		\begin{tabular}{l|l|l|ll}
			\hline
			\multirow{2}{*}{System} & \multicolumn{2}{l|}{Dataset 1} & \multicolumn{2}{l}{Dataset 2} \\ \cline{2-5} 
			& \begin{tabular}[c]{@{}l@{}}UBM\\ mixtures\end{tabular} & \begin{tabular}[c]{@{}l@{}}i-vector\\ dim\end{tabular} & \multicolumn{1}{l|}{\begin{tabular}[c]{@{}l@{}}UBM\\ mixtures\end{tabular}} & \begin{tabular}[c]{@{}l@{}}i-vector\\ dim\end{tabular} \\ \hline
			UBM-GMM & $128$ & - & \multicolumn{1}{l|}{$128$} & - \\ \hline
			baseline-i-vector & $64$ & $160$ & \multicolumn{1}{l|}{$64$} & $160$ \\ \hline
			modified-i-vector & $7$ & $160$ & \multicolumn{1}{l|}{$8$} & $160$ \\ \hline
		\end{tabular}%
	}
\end{table}

For dataset 1,  the $ix$-vector system gives the best performance, followed by the modified-$i$-vector system. For dataset 2, the $ix$-vector system gave the best accuracy, followed by the CNN-RNN based approach. However, in terms of EER, for dataset 2, CNN-RNN system has performed better.  We highlight that the proposed E-M based modified-$i$-vector gives comparable results to DNN based techniques even on a dataset with $920$ subjects. Further, the $ix$-vector system that combines embeddings from the modified-$i$-vector and modified-$x$-vector system, significantly improves upon the performance over individual systems. This result shows that the combined representation system is able to generalize better for test data. For both datasets, the proposed systems achieve state-of-the-art performances with accuracy as the evaluation metric. 

For dataset 2, the best-obtained accuracy is only $35.9\%$. We found that EEG recordings from many sessions in dataset 2 were still noisy and abnormal on manual observation.  However, we highlight that this is the highest number of subjects used for studying EEG biometrics  (see Section~\ref{sec:ds2_results}).

The proposed subspace systems extract embeddings from multi-channel EEG, which can then be used for recognizing individuals. In Section~\ref{sec:task_effect}~and~\ref{sec:unseen}, we explore the ability of these embeddings to scale for unseen tasks and individuals during training.

The embeddings from the proposed systems can be obtained for EEG segments of any length irrespective of the segments size used for training. In Table~\ref{tab:varying_seg}, we further evaluate the proposed models by  re-dividing the test data into EEG segments of duration $30$s and $60$s, in addition to $15$s.  It can be observed that, for all the proposed models, the accuracy increases as the duration of EEG segments increases. This shows that the effectiveness of the generated embeddings increases with the duration of the EEG segment.

\begin{table}[h!t]
	\centering
	\caption{Performances of the proposed system for different lengths of EEG segments.}
	\label{tab:varying_seg}
	\renewcommand{\arraystretch}{1.3}
	\resizebox{0.40\textwidth}{!}{%
		\begin{tabular}{c|c|c|c|c|c|c}
			\hline
			\multirow{2}{*}{\textbf{Systems}} & \multicolumn{3}{c|}{\textbf{Accuracy ($\%$)}}              & \multicolumn{3}{c}{\textbf{EER($\%$)}}                  \\ \cline{2-7} 
			& $15$s           & $30$s           & $60$s           & $15$s           & $30$s           & $60$s          \\ \hline 
			\multicolumn{7}{c}{\textbf{Dataset 1}} \\ \hline
			modified-$i$-vector                 & $85.1$ & $90.1$ & $93.0$ & $5.81$ & $4.32$ & $2.84$ \\ \hline
			modified-$x$-vector                 & $ 76.8 $          & $ 82.2 $          & $ 84.0 $          & $ 8.16 $          & $ 6.31 $          & $ 5.84 $         \\ \hline
			$ix$-vector                 & $ \mathbf{86.4} $          & $ \mathbf{90.8} $          & $ \mathbf{93.39} $          & $ \mathbf{5.02} $          & $ \mathbf{3.59} $          & $ \mathbf{2.59} $         \\ \hline
			\multicolumn{7}{c}{\textbf{Dataset 2}} \\ \hline
			modified-$i$-vector                 & $30.0$ & $ 37.7 $ & $42.8$ & $16.5$ & $ 14.5 $ & $13.3$ \\ \hline
			modified-$x$-vector                 & $ 27.2 $          & $ 32.8 $          & $ 36.4 $          & $ 16.6 $          & $ 15.0 $          & $ 14.0 $         \\ \hline
			$ix$-vector                 & $ \mathbf{35.9} $          & $ \mathbf{42.9} $          & $ \mathbf{47.0} $          & $ \mathbf{14.2} $          & $ \mathbf{12.6} $          & $ \mathbf{11.7} $         \\ \hline
	\end{tabular}}
\end{table}

\subsection{Evidence of task-independent person-specific signatures in EEG using proposed subspace systems}
\label{sec:task_effect}
In this experiment, the task-independent nature of the EEG biometric-signatures are tested in two steps as given below:
\begin{enumerate}
	\item The subspace embeddings are tested for their ability to normalize the variance across tasks seen during subspace training. 
	
	\item The same is tested for tasks unseen during training. 
\end{enumerate}

First, the subspace systems (modified-$i$-vector, modified-$x$-vector, and $ix$-vector) trained in Section~\ref{sec:result_1_comp} using all the tasks (Table~\ref{tab:2:experiment_protocols}) is used. Keeping EEG data of a particular task in Table~\ref{tab:2:experiment_protocols} for testing, the reference subspace vector is computed using other tasks performed on different sessions. Hence this experiment tests the task-independent nature of the vector embedding for a known task during subspace training. Later, the subspaces (modified-$i$-vector, modified-$x$-vector, and $ix$-vector) are trained again by leaving out a task from Table~\ref{tab:2:experiment_protocols} and using them only for testing. The results of both these studies of testing left-out tasks are given in Table~\ref{tab:left_out_task} using EEG segments of $15$s for evaluation.

\begin{table}[h!t]
	\centering
	\renewcommand{\arraystretch}{1.3}
	\caption{Accuracy (\%) and EER (\%) of tasks left-out for testing only. (System 1 : modified-$i$-vector; System 2 : modified-$x$-vector; System 3 : $ix$-vector)}
	\label{tab:left_out_task}
	\resizebox{0.48\textwidth}{!}{%
		\begin{tabular}{c|c|c||cc|cc}
			\hline
			\multirow{2}{*}{Left-out task for test} & \multirow{2}{*}{\begin{tabular}[c]{@{}l@{}}No. of\\ Subjects\end{tabular}} & \multirow{2}{*}{System} & \multicolumn{2}{c|}{\begin{tabular}[c]{@{}c@{}}Subspace\\ trained with \\ all tasks \\ (Case-1) \end{tabular}} & \multicolumn{2}{c}{\begin{tabular}[c]{@{}c@{}}Subspace \\ trained without \\ test task \\ (Case-2)\end{tabular}} \\ \cline{4-7} 
			&  &  & ACC & EER & ACC & EER \\ \hline \hline
			\multirow{3}{*}{Odd Ball Classic} & \multirow{3}{*}{ $ 13 $ } & 1 & $ 97.0 $ & $ 2.58 $ & $ 91.8 $ & $5.79$ \\ 
			&  & 2 & $ 89.1 $ & $ 3.93 $ & $83.0$ & $6.97$ \\ 
			&  & 3 & $ 96.6 $ & $ 1.71 $ & $ 94.5 $ & $ 4.07 $ \\ \hline
			\multirow{3}{*}{Odd Ball Stereo} & \multirow{3}{*}{ $ 12 $ } & 1 & $ 90.7 $ & $ 4.85 $ & $ 87.6 $ & $ 6.29 $ \\ 
			&  & 2 & $ 94.6 $ & $ 2.69 $ & $ 85.3 $ & $ 7.24 $ \\ 
			&  & 3 & $ 96.9 $ & $ 2.81 $ & $ 95.3 $ & $ 5.72 $ \\ \hline
			\multirow{3}{*}{\begin{tabular}[c]{@{}c@{}}Imagining \\ Binary Answers\end{tabular}} & \multirow{3}{*}{ $ 7 $ } & 1 & $ 96.9 $ & $ 2.93 $ & $94.0$ & $5.51$ \\ 
			&  & 2 & $ 99.1 $ & $ 1.62 $ & $ 84.1 $ & $ 12.7 $ \\ 
			&  & 3 & $ 99.3 $ & $ 1.39 $ & $ 96.4 $ & $ 5.36 $ \\ \hline
			\multirow{3}{*}{\begin{tabular}[c]{@{}c@{}}Semantically\\ Opposite Words\end{tabular}} & \multirow{3}{*}{ $ 4 $ } & 1 & $ 97.1 $ & $4.33$ & $88.8$ & $12.9$ \\ 
			&  & 2 & $ 95.1 $ & $ 6.64 $ & $ 82.1 $ & $ 17.1 $ \\ 
			&  & 3 & $ 99.5 $ & $ 0.91 $ & $ 88.4 $ & $ 11.3 $ \\ \hline
			\multirow{3}{*}{\begin{tabular}[c]{@{}c@{}}Familiar and \\ Unfamiliar Words\end{tabular}} & \multirow{3}{*}{ $ 6 $ } & 1 & $ 97.1 $ & $ 1.92 $ & $96.7$ & $4.36$ \\ 
			&  & 2 & $ 100 $ & $ 0.54 $ & $95.1$ &  $4.48$ \\ 
			&  & 3 & $  100 $ & $ 0.38 $ & $ 98.3 $ & $ 2.38 $ \\ \hline
			\multirow{3}{*}{\begin{tabular}[c]{@{}c@{}}Proper and \\ Improper Sentences\end{tabular}} & \multirow{3}{*}{ $ 8 $ } & 1 & $ 98.4 $ & $ 2.70 $ & $97.4$ & $2.68$ \\ 
			&  & 2 & $ 100 $ & $ 0 $ & $95.9$ & $ 2.53 $ \\ 
			&  & 3 & $ 100 $ & $ 0.06 $ & $ 98.4 $ & $ 1.95 $ \\ \hline
			\multirow{3}{*}{\begin{tabular}[c]{@{}c@{}}Motor and Mental\\ Imaginary\end{tabular}} & \multirow{3}{*}{$ 6 $} & 1 & $98.0$ & $ 2.04 $ & $96.5$ & $4.57$ \\ 
			&  & 2 & $ 98.0 $ & $ 2.9 $ & $ 95.3 $ & $ 5.56 $ \\ 
			&  & 3 & $ 100  $ & $ 1.44 $ & $ 96.92 $ & $ 3.77 $ \\ \hline
			\multirow{3}{*}{Passive Audio} & \multirow{3}{*}{ $17$ } & 1 & $ 82.5 $ & $7.42$ & $77.4$ & $11.7$ \\ 
			&  & 2 & $ 87.6 $ & $ 5.26 $ & $71.5$ & $7.06$ \\ 
			&  & 3 & $ 88.4  $ & $4.96 $ & $80.4$ & $7.90 $ \\ \hline
			\multirow{3}{*}{\begin{tabular}[c]{@{}c@{}}Passive Audio \\ Stereo\end{tabular}} & \multirow{3}{*}{ $11$ } & 1 & $89.8$ & $5.13$ & $ 91.0 $ & $ 4.94 $ \\ 
			&  & 2 & $ 89.3 $ & $ 2.69 $ & $ 89.8$  & $ 3.98 $  \\ 
			&  & 3 & $ 89.8 $ & $ 3.06 $ & $ 94.3 $ & $ 3.90 $ \\ \hline
			\multirow{3}{*}{Odd Ball Visual} & \multirow{3}{*}{ $ 6 $ } & 1 & $78.7$ & $13.7$ & $83.8$ &  $ 10.5 $ \\ 
			&  & 2 & $ 77.7 $  & $ 9.39 $  & $ 76.7 $  & $ 12.8 $ \\ 
			&  & 3 & $ 79.7 $ & $ 10.6 $ & $ 80.0 $ & $ 11.7 $ \\ \hline
			\multirow{3}{*}{\begin{tabular}[c]{@{}c@{}}Steady State Visually\\ Evoked Potential\end{tabular}} & \multirow{3}{*}{ $ 12 $ } & 1 & $80.1$ & $9.06$ & $ 69.3 $ & $16.4$ \\ 
			&  & 2 & $ 70.2 $  & $ 11.4 $ & $ 56.7 $ & $ 17.4 $ \\
			&  & 3 & $ 80.6 $ & $ 9.19 $ & $ 63.9 $ & $ 15.8 $ \\ \hline
			\multirow{3}{*}{Passive Audio-Visual} & \multirow{3}{*}{ $ 12 $ } & 1 & $87.3$ & $7.45$ & $ 78.2 $ & $ 10.5 $ \\ 
			&  & 2 & $ 72.3 $ & $ 14.7 $ & $ 57.1 $ & $21.0$ \\ 
			&  & 3 & $ 87.8 $ & $ 7.9 $ & $ 81.9 $ & $ 11.5 $ \\ \hline
		\end{tabular}%
	}
\end{table}

From Table~\ref{tab:left_out_task}, it can be seen that the results are slightly different for different tasks. However, for all tasks, the results are significantly high (accuracy  $\approx\ge78\%$ and EER $\approx\le10\%$ for modified-$i$-vector and $ix$-vector systems).  This result was obtained with a simple cosine similarity classifier using reference vectors that used no EEG data from the task and session under test. The results of this experiment show that, when trained with all tasks, the proposed approach can account for task and session related variance in the EEG data. When the task under test is also excluded from subspace training, the performance reduces slightly for most of the tasks. However, the results are still promising, showing that the proposed approach can extract person-specific signatures even for tasks and sessions  not seen during training.

To further test the task-independence of proposed systems, the tasks in Table~\ref{tab:2:experiment_protocols} were combined into data collected with the open and closed-eye conditions. It is well-known that the activation and inactivation of the visual cortex by open/closed eye conditions have a significant impact on the brain activation patterns \cite{marx2004eyes}. Using EEG, the  open/closed eye state can be classified with $\approx$$97\%$ accuracy \cite{rosler2013first}.  We again repeat the previous experiment but only using two conditions (the open and closed-eye conditions) rather than all tasks from Table~\ref{tab:2:experiment_protocols}. Out of all the $30$ subjects from dataset $1$ only $10$ subjects have recordings with both open and closed-eye conditions during all the sessions. Only the data from those $10$ subjects are used in this experiment. First, we use the subspace trained using all the tasks from Section~\ref{sec:result_1_comp} and extract the reference vector without using the data from the condition used for testing. Later, we repeat the experiment using a subspace that was trained without the condition reserved for testing. The result of these experiments across the open and closed-eye conditions are given in Table~\ref{tab:6:mismatch-testing} using EEG segments of $15$s for all the evaluation.  

\begin{table}[h!t]
	\centering
	
	\caption{Accuracy (\%) and EER (\%) of  open/closed eye condition left-out for testing only. (System 1 : modified-$i$-vector; System 2 : modified-$x$-vector; System 3 : $ix$-vector)}
	\label{tab:6:mismatch-testing}
	\renewcommand{\arraystretch}{1.3}
	\resizebox{0.48\textwidth}{!}{%
		\begin{tabular}{c|c|c||cc|cc}
			\hline
			\multirow{2}{*}{Left-out condition for test} & \multirow{2}{*}{\begin{tabular}[c]{@{}l@{}}No. of\\ Subjects\end{tabular}} & \multirow{2}{*}{System} & \multicolumn{2}{c|}{\begin{tabular}[c]{@{}c@{}}Subspace\\ trained with \\ all tasks \\ (case-1) \end{tabular}} & \multicolumn{2}{c}{\begin{tabular}[c]{@{}c@{}}Subspace \\ trained without \\ test condition \\ (case-2) \end{tabular}} \\ \cline{4-7} 
			&  &  & Acc & EER & ACC & EER \\ \hline \hline
			\multirow{3}{*}{ Closed Eye Condition} & \multirow{3}{*}{$ 10 $} & 1 & $82.6$ & $7.50$ & $ 35.8 $ & $ 29.8 $ \\ 
			&  & 2 & $86.8$ & $ 4.18 $ & $ 39.3 $ &  $ 32.4 $ \\ 
			&  & 3 & $ 88.3 $ & $ 4.34 $ & $37.0 $ & $ 29.9 $ \\ \hline
			\multirow{3}{*}{ Open Eye Condition} & \multirow{3}{*}{$ 10 $} & 1 & $85.7$ & $7.59$ & $67.2$ & $  14.84 $ \\ 
			&  & 2 & $80.0$ & $7.47$ & $52.3$ & $19.3$ \\ 
			&  & 3 & $ 86.6 $ & $ 6.02 $ & $ 60.4 $ & $ 13.3 $ \\ \hline
			
		\end{tabular}%
	}
\end{table}

The results in Table~\ref{tab:6:mismatch-testing}, show that even when the subspace representation obtained from  the open/closed-eye condition is tested against the reference vector formed using the opposite  condition, the simple cosine similarity measure is able to recognize the individuals with an EER less than $\approx7\%$. This suggests that the proposed subspace techniques have the ability to model the variations across major changes in the underlying circuit, such as activation and inactivation of the vision system. In Table~\ref{tab:6:mismatch-testing}, when the subspace is trained without the test condition, the performance degrades. Especially when eye closed condition is retained only for test, the EER reduces to $\approx35\%$. Although this result is significantly higher than chance, it clearly shows that the proposed subspace may not scale if the underlying circuit generating the EEG changes significantly. Further, the results also show that, if the same variability in tasks are seen in the  training data, the proposed approach is able to extract task-independent features.

\subsection{The subspace systems generalize to unseen subjects}
\label{sec:unseen}
Training of the entire subspace for every new user to be enrolled in a biometric system can be a tedious task. When trained using many subjects, the subspace should represent biometric signatures independent of specific individuals used during training. This hypothesis is well-established in the speaker recognition literature. The speakers used for evaluation are seldom used during subspace training.  Taking this idea forward, we test the subspace system performance for subjects not seen during training.

A random $20\%$ of the subjects from both datasets 1 and 2 were selected for evaluation in this experiment. The performance for this $20\%$ of the random subjects is shown under two cases. In Case 1, all the subjects are used for training, and evaluation is done only on the selected $20\%$  subjects. In Case 2, we retain the selected $20\%$ of the subjects for only testing and repeat the entire subspace training procedure using only the remaining $80\%$ of the subjects. The results for Case 1 and 2 are compared in Table~\ref{tab:7:unseen_subject} for both datasets using EEG segments of 15 seconds. It can be observed that the results degrade when the subjects are not included in the training pipeline. However, with dataset 2, this degradation in performance is smaller compared to dataset 1. 
\begin{table}[h!t]
	
	\renewcommand{\arraystretch}{1.3}
	\centering
	\caption{Accuracy (\%) and EER (\%) when subspace is trained with all subjects including the subjects under test (Case 1) vs when the subspace is trained without the subjects under test (Case 2).}
	\resizebox{0.38\textwidth}{!}{
		\begin{tabular}{clc|c|c|c|c}
			\cline{4-7}
			\multicolumn{1}{l}{} & \multicolumn{1}{l}{} & \multicolumn{1}{l}{} & \multicolumn{2}{c|}{Case 1} & \multicolumn{2}{c}{Case 2}                                         \\ \cline{4-7}
			&              &     \multicolumn{1}{l}{}     &    ACC           & EER             & ACC            & EER                                                   \\ \cline{1-7} 
			\multicolumn{1}{l|}{\multirow{12}{*}{\begin{turn}{-90}Dataset 1\end{turn}}} & \multicolumn{1}{c|}{\multirow{4}{*}{\begin{tabular}[c]{@{}c@{}}Modified-\\ $i$-vector\end{tabular}}} &Split-1            & $ 92.3 $ & $ 6.85 $ & $ 70.3 $ & $ 14.9 $   \\ \cline{3-7}
			\multicolumn{1}{c|}{} & \multicolumn{1}{l|}{}  & Split-2 & $ 95.4 $ & $ 4.97 $ & $ 90.3 $ & $ 13.6 $                  \\ \cline{3-7}
			\multicolumn{1}{c|}{} & \multicolumn{1}{l|}{}  & Split-3 & $ 89.4 $ & $ 10.3 $ & $ 80.5 $ & $ 14.5 $                   \\ \cline{3-7}
			\multicolumn{1}{c|}{} & \multicolumn{1}{l|}{}  &Avg      & $ 92.4 $ & $ 7.37 $ & $ 80.4 $ & $ 14.3 $                   \\ \cline{2-7}
			\multicolumn{1}{c|}{} & \multicolumn{1}{c|}{\multirow{4}{*}{\begin{tabular}[c]{@{}c@{}}Modified-\\ $x$-vector\end{tabular}}} &Split-1             & $ 92.3 $ & $ 4.87 $ & $ 64.3 $ & $ 19.4 $                    \\ \cline{3-7}
			\multicolumn{1}{c|}{} & \multicolumn{1}{l|}{} & Split-2 & $ 88.1 $ & $ 8.72 $ & $ 63.75 $ & $ 22.8 $                 \\ \cline{3-7}
			\multicolumn{1}{c|}{} & \multicolumn{1}{l|}{} & Split-3 & $ 91.1 $ & $ 6.91 $ & $ 88.9 $ & $ 11.5 $                    \\ \cline{3-7}
			\multicolumn{1}{c|}{}& \multicolumn{1}{l|}{} &Avg       & $ 90.8 $ & $ 6.83 $ & $ 72.3 $ & $ 17.9 $                    \\  \cline{2-7} 
			\multicolumn{1}{c|}{} &
			\multicolumn{1}{c|}{\multirow{4}{*}{\begin{tabular}[c]{@{}c@{}}$ix$-vector\end{tabular}}} &Split-1             & $95.2 $ & $ 3.96 $ & $ 76.7 $ & $ 13.3 $                    \\ \cline{3-7}
			\multicolumn{1}{c|}{} & \multicolumn{1}{l|}{} & Split-2 & $ 98.1 $ & $ 3.84 $ & $81.6  $ & $16.1 $                 \\ \cline{3-7}
			\multicolumn{1}{c|}{} & \multicolumn{1}{l|}{} & Split-3 & $ 95.9 $ & $ 4.42 $ & $ 92.4 $ & $  7.83 $                    \\ \cline{3-7}
			\multicolumn{1}{c|}{}& \multicolumn{1}{l|}{} &Avg       & $ 96.4 $ & $ 4.08 $ & $ 83.6  $ & $ 12.4  $                    \\  \cline{1-7} \hline
			
			\multicolumn{1}{l|}{\multirow{12}{*}{\begin{turn}{-90}Dataset 2\end{turn}}} &  \multicolumn{1}{c|}{\multirow{4}{*}{\begin{tabular}[c]{@{}c@{}}Modified-\\ $i$-vector\end{tabular}}} & Split-1          & $ 42.4 $ & $ 14.2 $ & $ 35.8 $ & $ 16.1 $ \\ \cline{3-7}
			\multicolumn{1}{c|}{}& \multicolumn{1}{l|}{}  & Split-2 & $ 42.4 $ & $ 14.2 $ & $ 35.8 $ & $ 16.1 $               \\ \cline{3-7}
			\multicolumn{1}{c|}{} & \multicolumn{1}{l|}{} & Split-3 & $ 36.9 $ & $ 17.7 $ & $ 34.3 $ & $ 18.7 $                  \\ \cline{3-7}
			\multicolumn{1}{c|}{} & \multicolumn{1}{l|}{} & Avg     & $ 40.5 $ & $ 15.3 $ & $ 35.3 $ & $ 17.0  $                  \\ \cline{2-7}
			\multicolumn{1}{l|}{} &\multicolumn{1}{c|}{\multirow{4}{*}{\begin{tabular}[c]{@{}c@{}}Modified-\\ $x$-vector\end{tabular}}} & Split-1 & $ 40.3 $ & $ 16.2 $ & $ 22.5 $ & $ 23.5 $                    \\ \cline{3-7}
			\multicolumn{1}{c|}{} & \multicolumn{1}{l|}{} & Split-2               & $ 36.4 $ & $ 18.2 $ & $ 24.1 $ & $ 23.6 $                   \\ \cline{3-7}
			\multicolumn{1}{c|}{}  & \multicolumn{1}{l|}{}                                                                              & Split-3               & $ 36.6 $ & $ 17.6 $ & $ 33.3 $ & $ 19.3 $                   \\ \cline{3-7}
			\multicolumn{1}{c|}{} & \multicolumn{1}{l|}{} & Avg & $ 37.7 $ & $ 17.3 $                & $ 26.7  $ & $ 22.1 $                   \\ \cline{2-7}
			\multicolumn{1}{c|}{} &
			\multicolumn{1}{c|}{\multirow{4}{*}{\begin{tabular}[c]{@{}c@{}}$ix$-vector\end{tabular}}} &Split-1             								& $43.0 $ & $ 14.8 $ & $ 41.1 $ & $ 15.3 $                    \\ \cline{3-7}
			\multicolumn{1}{c|}{} & \multicolumn{1}{l|}{} & Split-2 & $ 49.4 $ & $ 12.6 $ & $ 37.3 $ & $ 15.3 $                 \\ \cline{3-7}
			\multicolumn{1}{c|}{} & \multicolumn{1}{l|}{} & Split-3 & $ 44.3 $ & $ 15.3 $ & $ 37.7 $ & $ 17.3 $                    \\ \cline{3-7}
			\multicolumn{1}{c|}{}& \multicolumn{1}{l|}{} &Avg       & $ 45.6 $ & $ 14.2 $ & $ 38.7 $ & $ 16.0 $                    \\  \cline{1-7} \hline
	\end{tabular}}
	\label{tab:7:unseen_subject}
\end{table}

\subsection{Channels needed for effective estimation on the subspace signatures}
\label{sec:channel_effect}
All the results reported in other sections of this paper have used only the $9$ channels mentioned in Section~\ref{sec:Experimental_Setup}. This section empirically explores various spatial subsampling methods to analyze the set of channels that perform better for task-independent EEG person recognition. All results in this section using only dataset 1 with $30$ subjects and $128$ EEG channels.

\begin{figure*}[h!t]
	\centering
	\includegraphics[width=1\textwidth]{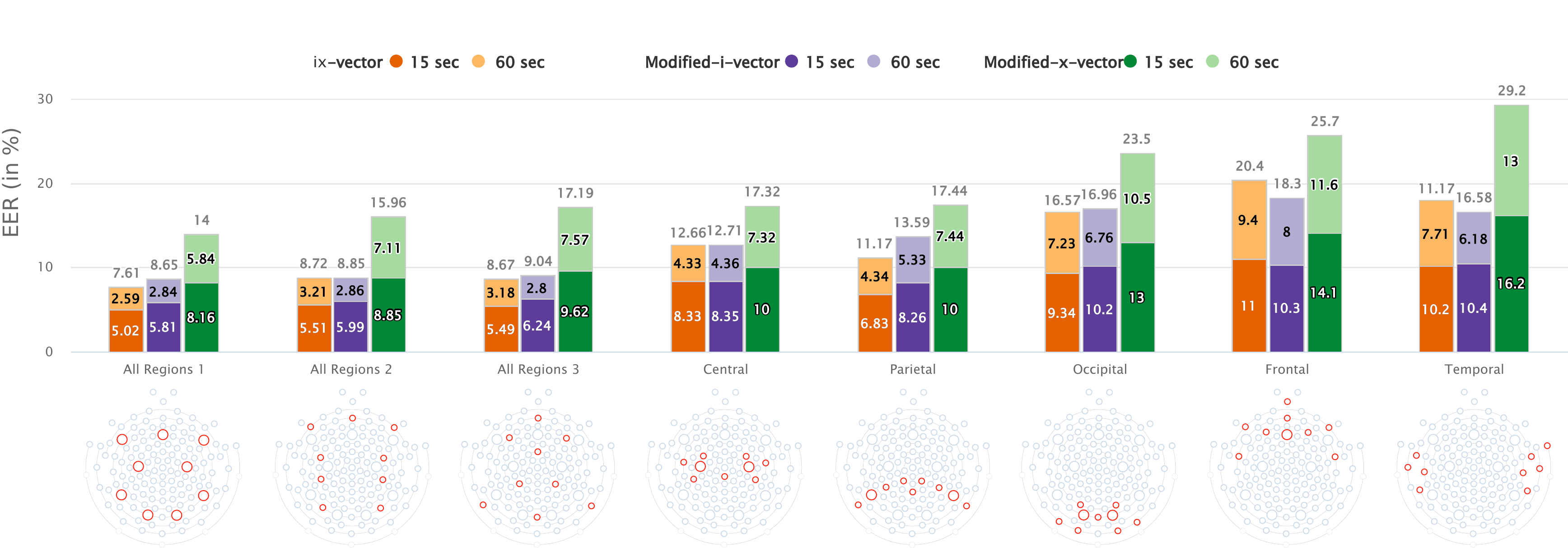}
	\caption{Performance of sensors sampled from different regions of the brain. The channel map below highlights the sensors selected for each case. The sensor with larger radius represents the 9 primary channels used in all the other experiments.}
	\label{fig:5:area_wise}
\end{figure*}

Given a particular number of sensors, we first explore different possible ways to sample them from the available $128$ channels. Accordingly, in Figure~\ref{fig:5:area_wise},  nine channels are sampled locally from different regions of the brain, namely, Frontal, Central, Parietal, Temporal, and Occipital lobes. In addition, three combinations of sensors are chosen such that they cover all the regions equally. The $ix$-vector and modified-$i$-vector systems were observed to recognize subjects with much better EERs when sensors are sampled from all the regions  rather than locally from a particular region. However, among the different areas of the brain, the  Central region is observed to give better recognition, followed by the Parietal region. Nevertheless,  by showing consistently good EERs for three different selections, Figure~\ref{fig:5:area_wise} shows that sampling sensors from across the entire scalp yield a better result for task-independent person recognition.

\begin{figure*}[h!t]
	\centering
	\includegraphics[width=1\textwidth]{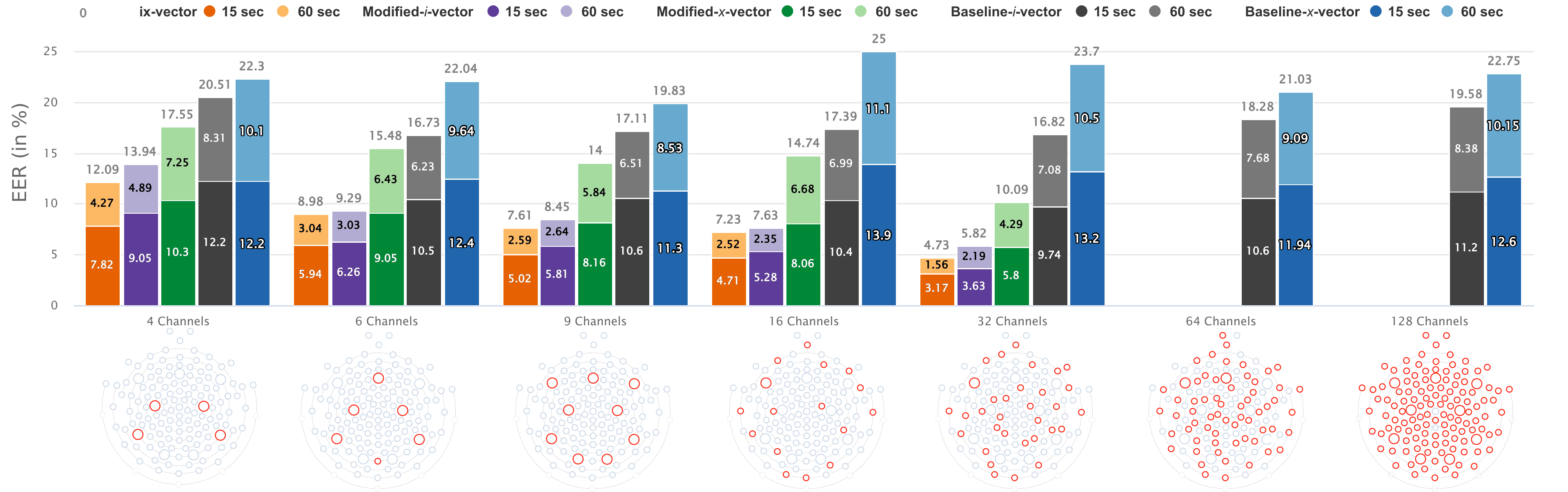}
	\caption{Performance of systems built with different number of Channels. The channel map below highlights the sensors selected for each case. The sensor with larger radius represents the 9 primary channels used in all the other experiments.}
	\label{fig:6:number_of_channel_testing}
\end{figure*}

In Figure~\ref{fig:6:number_of_channel_testing}, we analyze the number of channels required for EEG biometrics using channel subsets of different sizes from the available $128$ sensors. The sensors for systems with $16$, $32$, and $64$ channels are sampled by incrementing the channel numbers (given in Figure~\ref{fig:2:128channel}) by $8$, $4$, and $2$, respectively. This selection ensured that the entire scalp was covered. The EER of the systems using a larger number of sensors are compared with the $9$ channel systems used in other sections of this paper. The results for the proposed systems with channels $\ge 64$ are not shown because higher the number of channels, greater the data needed to train the system owing to the concatenation of statistics across channels (Section~\ref{sec:Methods}). In Figure~\ref{fig:6:number_of_channel_testing}, observe that the modified-$i$-vector system using just $9$ channels achieves an EER of $5.81\%$. Whereas, using all the $128$ channels, the baseline-$i$-vector system achieves a performance of $11\%$ EER.  These results suggest that not all channels are required for subject identification. Nevertheless, it is important to ensure that the sensors are spatially apart and cover the entire scalp. Further, using $32$ channels, the performance of the $ix$-vector and modified-$i$-vector systems are observed to increase gracefully.

To study the performance of systems with fewer than $9$ EEG sensors, we also analyze systems with $6$ and $4$ channels in Figure~\ref{fig:6:number_of_channel_testing}. While selecting sensors for the $6$ and $4$ channel system, higher importance was given to Central and Parietal regions as they were observed to provide better recognition in Figure~\ref{fig:5:area_wise}. It is interesting that the system with just $4$ channels is found to give slightly better recognition than baseline systems using all the $128$ channels. Adding more channels need more  data to train the systems. Further, as more channels are added, this improvement in performance degrades gradually. This is a significant result as it indicates that additional channels only lead to redundant information and increased dimensionality requiring more training data.

\subsection{Proposed modifications for the $i$-vector and the $x$-vector systems vs. naive early and late fusion techniques on baseline models}
\label{sec:modifications_adv}
In Section~\ref{sec:result_1_comp}, it was observed that explicit modeling of the EEG sensors by the proposed approaches gave a significant improvement in performance over the baseline versions adopted from speaker recognition. In the literature of EEG biometrics, information from multiple channels is handled by either concatenating the input feature vector or by performing voting/score fusion on channel-specific models (Table~\ref{tab:1:Literature_Survey}). In this section, we explicitly model the EEG channels in baseline versions of the $i$-vector and the $x$-vector system by feature concatenation and score fusion. Table~\ref{tab:4:chan_i_Vec} compares performance of proposed $i$-vector systems with the early and late fusion versions.
Table~\ref{tab:5:chan-x-vec} presents a similar comparison for $x$-vector based systems. Similar to  Section~\ref{sec:result_1_comp}, the experimental setup detailed in Section~\ref{sec:Experimental_Setup} was followed and the hyper-parameters for all systems were fine-tuned using validation data.

\begin{table}[h!t]
	\centering
	\renewcommand{\arraystretch}{1.3}
	\caption{Performance comparison of different ways of modeling EEG channels in $i$-vector framework}
	\resizebox{0.38\textwidth}{!}{%
	\begin{tabular}{c|l|c|c|c|c}
		\hline
		\multirow{2}{*}{\textbf{Systems}}                                                                   & \multicolumn{1}{c|}{\multirow{2}{*}{\textbf{\begin{tabular}[c]{@{}c@{}}Channel\\ Information\end{tabular}}}} & \multicolumn{2}{c|}{\textbf{Accuracy}}        & \multicolumn{2}{c}{\textbf{EER}}             \\ \cline{3-6} 
		& \multicolumn{1}{c|}{}                                                                                        & $15$s                 & $60$s                 & $15$s                 & $60$s                 \\ \hline
		\multicolumn{6}{c}{\textbf{Dataset 1}} \\ \hline
		\multicolumn{1}{l|}{\multirow{3}{*}{\begin{tabular}[c]{@{}l@{}}baseline-\\ $i$-vector\end{tabular}}} & -             & $ 70.5 $ & $ 84.5 $ & $ 10.6 $ & $ 6.51 $ \\ \cline{2-6} 
		\multicolumn{1}{l|}{}                                                                              & Concatenation    & $ 79.5 $ & $ 86.4 $ & $ 7.41 $ & $ 4.74 $ \\ \cline{2-6} 
		\multicolumn{1}{l|}{}                                                                              & Score-Fusion     &$ 62.7 $ & $ 67.1 $ & $ 15.2 $ & $ 13.6 $ \\ \hline
		\begin{tabular}[c]{@{}c@{}}modified-\\ $i$-vector\end{tabular}                                        & \begin{tabular}[c]{@{}l@{}}Statistics\\ Concatenation\end{tabular}                                           & $\mathbf{85.1}$       & $\mathbf{93.0}$       & $\mathbf{5.81}$       & $\mathbf{2.84}$        \\ \hline
		\multicolumn{6}{c}{\textbf{Dataset 2}} \\ \hline
		\multirow{3}{*}{\begin{tabular}[c]{@{}c@{}}baseline-\\ $i$-vector\end{tabular}}                       & -                                                                                                            & $ 9.02 $                & $ 16.4 $                & $ 26.8 $                & $ 22.2 $                \\ \cline{2-6} 
		& Concatenation                                                                                                & $ 27.0 $ & $ 38.2 $ & $ 16.6 $ & $ 14.1 $  \\ \cline{2-6} 
		& Score-Fusion                                                                                                 & $ 13.6 $ & $ 19.4 $ & $ 20.6 $ & $ 17.6 $ \\ \hline
		\begin{tabular}[c]{@{}c@{}}modified-\\ $i$-vector\end{tabular}                                        & \begin{tabular}[c]{@{}l@{}}Statistics\\ Concatenation\end{tabular}                                           & $\mathbf{30.0}$                & $\mathbf{42.8}$                & $\mathbf{16.5}$                & $\mathbf{13.3}$                \\ \hline
	\end{tabular} }
	\label{tab:4:chan_i_Vec}
\end{table}

\begin{table}[h!t]
	\renewcommand{\arraystretch}{1.3}
	\caption{Performance comparison of different ways of modeling EEG channels in $x$-vector framework}
	\centering
	\resizebox{0.38\textwidth}{!}{%
	\begin{tabular}{c|l|c|c|c|c}
		\hline
		\multirow{2}{*}{\textbf{Systems}}                                                                   & \multicolumn{1}{c|}{\multirow{2}{*}{\textbf{\begin{tabular}[c]{@{}c@{}}Channel\\ Information\end{tabular}}}} & \multicolumn{2}{c|}{\textbf{Accuracy}}        & \multicolumn{2}{c}{\textbf{EER}}             \\ \cline{3-6} 
		& \multicolumn{1}{c|}{}                                                                                        & $15$s                 & $60$s                 & $15$s                 & $60$s                 \\ \hline
		\multicolumn{6}{c}{\textbf{Dataset 1}} \\ \hline
		\multicolumn{1}{l|}{\multirow{3}{*}{\begin{tabular}[c]{@{}l@{}}baseline-\\ $x$-vector\end{tabular}}} & -             & $ 67.1 $ & $ 74.7 $ & $ 11.3 $ & $ 8.53 $ \\ \cline{2-6} 
		\multicolumn{1}{l|}{}                                                                              & Concatenation    & $ 61.2 $ & $ 68.0 $ & $ 12.7 $ & $ 11.4 $ \\ \cline{2-6} 
		\multicolumn{1}{l|}{}                                                                              & Score-Fusion     &$ 73.7 $ & $ 80.5 $ & $ 10.8 $ & $ 9.26 $ \\ \hline
		\begin{tabular}[c]{@{}c@{}}modified-\\ $x$-vector\end{tabular}                                        & \begin{tabular}[c]{@{}l@{}}Statistics\\ Concatenation\end{tabular}                                           & $\mathbf{76.8}$       & $\mathbf{84.0}$       & $\mathbf{8.16}$       & $\mathbf{5.84}$        \\ \hline
		\multicolumn{6}{c}{\textbf{Dataset 2}} \\ \hline
		\multirow{3}{*}{\begin{tabular}[c]{@{}c@{}}baseline-\\ $x$-vector\end{tabular}}                       & -                                                                                                            & $ 3.64 $                & $ 5.01 $                & $ 32.8 $                & $ 29.7 $                \\ \cline{2-6} 
		& Concatenation                                                                                                & $ 9.036 $ & $ 11.8 $ & $ 27.9 $ & $ 24.5 $ \\ \cline{2-6} 
		& Score-Fusion                                                                                                 & $ 9.56 $ & $ 13.7 $ & $ 27.2 $ & $ 24.3 $ \\ \hline
		\begin{tabular}[c]{@{}c@{}}modified-\\ $x$-vector\end{tabular}                                        & \begin{tabular}[c]{@{}l@{}}Statistics\\ Concatenation\end{tabular}                                           & $\mathbf{27.2}$                & $\mathbf{36.4}$                & $\mathbf{16.6}$                & $\mathbf{14.0}$                \\ \hline
	\end{tabular} }
	\label{tab:5:chan-x-vec}
\end{table}

From Tables~\ref{tab:4:chan_i_Vec} and \ref{tab:5:chan-x-vec}, it can be seen that modified versions of the $i$-vector and the $x$-vector systems have outperformed the baseline versions that use explicit channel information through early concatenation or late fusion.  In the case of $i$-vector, feature concatenation gives a better result than score fusion. Owing to discriminative training, the $x$-vector systems trained on individual channels are better than $i$-vector trained on a single channel. Hence the $x$-vector model gives a better score fusion result compared to the baseline and concatenated approach. However, for both the $x$-vector and $i$-vector system, the proposed method is observed to give the best performance by concatenating statistics from various channels at an intermediate level of processing.

\section{Discussion}
\label{sec:Discussion}

\subsection{Task-independent EEG biometrics}
\label{sec:diss:task_effect}
The person-specific signatures in EEG were tested for task-independence in Section~\ref{sec:task_effect} using mismatched tasks for train and test. First, the task-independence was tested with subspace trained using all the data and with a mismatched task/condition for the cosine similarity back-end.  Later, the subspace system was also retrained without using the task/condition reserved for the test. For both these cases, in Tables~\ref{tab:left_out_task}~and~\ref{tab:6:mismatch-testing}, all the results are significantly above chance accuracy. This result shows that biometric information is present in EEG irrespective of the task and condition. From Tables~\ref{tab:left_out_task}~and~\ref{tab:6:mismatch-testing}, it is also evident that the subspace model generalizes better when the task/condition under test is also used for training the subspace.

\begin{figure*}[h!t]
	\centering
	\includegraphics[width=1\textwidth]{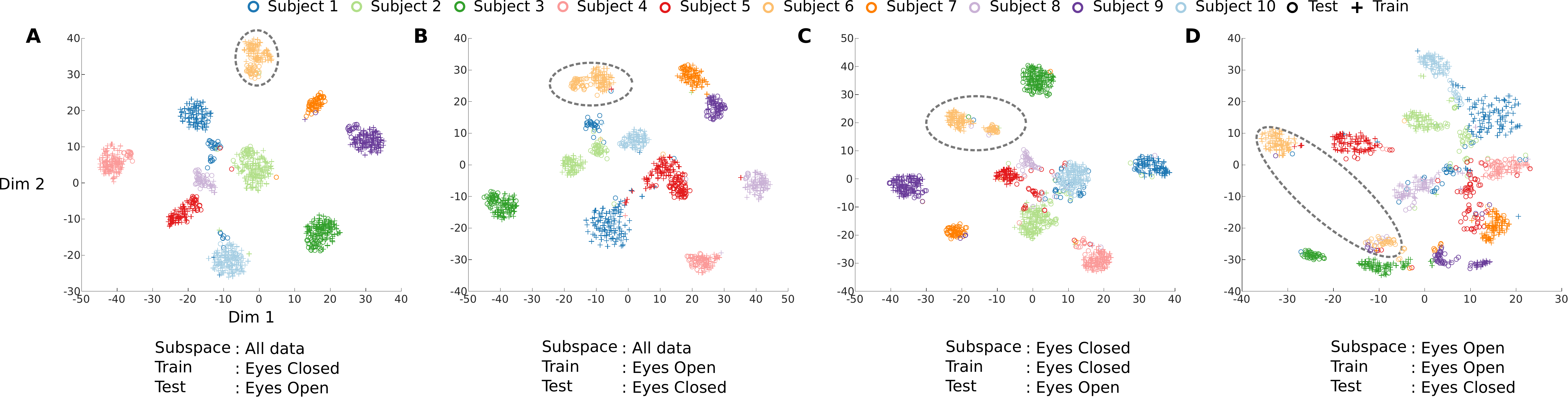}
	\caption{2-D visualization of modified-$i$-vector subspace for mismatched train and test conditions given in Table~\ref{tab:6:mismatch-testing}. All the training data are used to form a single reference vector for cosine similarity testing. }
	\label{fig:4:taskeffect}
\end{figure*}

In Table~\ref{tab:6:mismatch-testing}, the worst performance of EER  $\approx30\%$ was obtained when the subspace was trained using the closed eye condition and tested on the open eye condition. To further analyze this, the embeddings of the modified-$i$-vector system were projected to a two-dimensional space using t-distributed stochastic neighbor embedding (t-SNE)  \cite{maaten2008visualizing}. t-SNE is a non-linear dimension reduction technique such that it preserves the distance between the two points in the original space. t-SNE is applied to reduce the EEG signatures in the $i$-vector space to a visualizable $2D$ space. In Figure~\ref{fig:4:taskeffect}, the t-SNE plots were made for all four conditions given in Table~\ref{tab:6:mismatch-testing}. 

In Figure~\ref{fig:4:taskeffect}.~A~and~B, it can be seen that when the subspace is trained with all the data, the EEG segments from different conditions and sessions are located close to each other. The symbol ``+" denotes the training condition from which the reference vector is formed, and ``o" indicates the test condition. In Figure~\ref{fig:4:taskeffect}.~A~and~B, subject $6$ highlighted by a black circle, has two distinct clusters made of data from the train (``+") and test (``o") conditions close to each other. This shows the ability of the subspace to normalize the variance related to tasks and sessions. In Figure~\ref{fig:4:taskeffect}.~C~and~D, the distance between train and test clusters increases when data from test conditions are not used to train the subspace. It is well known that alpha oscillations are present in the occipital lobe when the eyes are closed. The subspace trained only using open eye condition is unaware of the dominant alpha oscillations in closed eye condition.  These alpha oscillations could be one of the reasons for poor task normalization in Figure~\ref{fig:4:taskeffect}.~D (subject 6, for example).
Hence, the subspace model cannot be expected to scale for a significant change in the underlying EEG  (such as alpha waves during closed eye or change in EEG due to a brain injury). However, when these conditions are included while training the subspace, the model is able to generalise even across such conditions. 

\subsection{Significance of the subspace systems}
\label{sec:performance_proposed_system}
The most important observation from Section~\ref{sec:result_1_comp} is that the proposed approaches, namely,  modified-$i$-vector and modified-$x$-vector system, give a significant improvement in performance over the baseline systems. Both the baseline-$i$-vector and the baseline-$x$-vector systems (Section~\ref{sec:Methods}) assume that the biometric information is present in the entire EEG signal. Hence these systems do not perform any sequence modeling. These subspace systems accumulate statistics across time in a higher dimension space and then project to a lower-dimensional space such that the biometric information is preserved.  The UBM-GMM system and baseline versions of the subspace systems do not model the data from different sensors explicitly.

The proposed modification suggests the pooling of data to create a common high-dimensional space for all channels. In the high-dimensional space, various statistics are estimated across time for each channel. These statistics are then concatenated across channels and reduced to a single vector in a subspace that enhances person-specific information. When this channel information is explicitly modeled by the modified version of the $i$-vector and the $x$-vector frameworks, a significant improvement in performance is observed in the respective systems. This observation is consistent with both accuracy and EER in Table~\ref{tab:all_result}. This improvement in the performance of the proposed subspace systems over the baseline systems demonstrates that the former can better model the person-specific signatures from the EEG signal independent of tasks.

Sections~\ref{sec:task_effect}~and~\ref{sec:unseen} tested the modified-$i$-vector and the modified-$x$-vector subspaces with tasks/conditions and subjects that were not used to train the subspace, respectively. The results presented in these experiments lead to interesting future works (See Section~\ref{sec:fut}). In addition to this, in the supplementary material (Section S1), we show preliminary results on longitudinal testing using the proposed models.

The proposed approach has a drawback that the same set of channels should be present during training and testing as compared to the baseline subspace systems. In addition, owing to the concatenation of statistics, these models require more data when the number of channels is increased. Nevertheless, Section~\ref{sec:channel_effect} shows that the modified-$i$-vector system with just $4$ channels outperforms the baseline-$i$-vector model using all the $128$ channels. This result is significant because using a larger number of channels is not feasible for building real-time biometric systems using EEG. In addition, we also show empirically that sampling electrodes from across the entire scalp gives a better EER than choosing the sensors locally from a particular region of the brain. In this regard, in the supplementary material (Section S2),  we also demonstrate each channel's contribution in the multichannel subspace embedding.

Finally, in Section~\ref{sec:modifications_adv}, we empirically show that the proposed technique of concatenating the statistics at an intermediate level of processing is better than simple early or late fusion applied the baseline models adopted as such from speaker recognition literature.

\subsection{Significance of combining embeddings from modified-$i$-vector and modified-$x$-vector}
This paper proposes a novel $ix$-vector system that concatenates the embeddings from modified-$i$-vector and modified-$x$-vector systems. Across both the datasets, in Table~\ref{tab:all_result} the $ix$-vector system is observed to perform better than other subspace systems and the baseline systems. In Case-1 of Table~\ref{tab:left_out_task}, observe that for few tasks the modified-$i$-vector system has given good performance and for the rest the modified-$x$-vector has given good results. However, for all the tasks, $ix$-vector provides better results than any of the individual systems.  Furthermore, for most of the results described in Sections~\ref{sec:result_1_comp}~to~\ref{sec:channel_effect}, the same trend is observed. This suggests that modified-$i$-vector and modified-$x$-vector capture complementary information, which is evident in the $ix$-vector system, where the performance is better than that of both $i$-vector and $x$-vector based systems. The only exception to this result is in the case when region-wise channels are chosen (Figure~\ref{fig:5:area_wise}).

\subsection{Results on dataset 2 with $920$ individuals}
\label{sec:ds2_results}
In Table~\ref{tab:all_result}, all the methods obtain accuracy only in the range of $5\%$-$35\%$  for dataset $2$. We believe there are two main factors for this drop in performance. Firstly, dataset $2$ with $920$ subjects is a much larger dataset with  multi-session recordings on which EEG person recognition is studied until now  (around $100$ subjects in Table~\ref{tab:1:Literature_Survey}). Secondly, unlike dataset 1 or any studies referred-to in Table~\ref{tab:1:Literature_Survey}, dataset $2$ was not collected in a controlled and confined environment with a predefined set of standard EEG biometrics elicitation protocols on $100\%$ healthy subjects. As mentioned in Section~\ref{sec:dataset2}, we have tried to remove recordings with abnormalities or seizures. However, the annotation used to filter the dataset were algorithmically generated \cite{Lopez2017} and hence are not expected to be $100\%$ accurate. Abnormalities such as seizure would be present in only a few sessions, and such signatures would have made the dataset more challenging. Contrasting this, if the abnormality is consistently present in all the sessions, this would have positively enhanced biometric results. Non-standardized tasks and data collecting environments, and varying EEG acquisition devices further added challenges to biometric performance in the dataset.  Owing to the factors mentioned above, finding a subspace that disentangles the intrinsic subject characteristics from the EEG signal in this dataset is hard.

Further experiments are required to decode the extent to which the clinical nature and a large number of subjects affects EEG biometric performance in dataset 2. The performance of EEG biometrics on a large number of subjects with multiple session recording is an unexplored area, mainly owing to the difficulty involved in data collection (unlike face or
voice recognition where $>1000$ individuals or more is typically used). The given result on dataset $2$ with $920$ subjects  further motivates the research of EEG biometrics with a large number of subjects and multiple sessions of recording, although the clinical nature influences the results. In the supplementary material (Section S3), we show further analysis on this dataset using $100$ and $500$ subjects.

\subsection{Limitations}
\label{sec:limitations}
\subsubsection{Dataset $1$}
 The limitation of dataset $1$ is that it does not have data from all or an equal number of subjects performing all the tasks. Hence, it is not possible to analyze the performance from Table~\ref{tab:left_out_task} across tasks and compare them. However, dataset $1$ has been collected using a wide range of tasks, making it a suitable candidate for task-independent EEG analysis. 
 \subsubsection{Dataset $2$}
 Dataset 2 has clinical nature due to the demographics of the involved patients and the usage of non-standardized tasks with clinical objectives. Further studies are required to carefully analyze the effect and the extent of clinical nature in this dataset and how it affects EEG biometrics. This paper uses dataset 2 only as a contrastive dataset to compare the performance of proposed approaches against the baseline systems (see Section~\ref{sec:ds2_results}).

\subsection{Future research directions}
\label{sec:fut}
 \subsubsection{Task-independent EEG biometrics}
 With different elicitation protocols being a primary focus in the literature, all the results discussed in this paper question the need for specific (constrained) elicitation protocols for studying biometric signatures. These results show that there can be significant person-specific signatures in any EEG being collected; and hence suggest the usage of task-independent EEG biometrics as a baseline for studying task-dependent EEG biometrics.
 
 \subsubsection{Proposed subspace}
 The person-specific signatures can negatively affect generalization across individuals when EEG is being used for building task rich Brain-Computer Interfaces (BCI). In the speech processing literature,  while training models for speech recognition, speaker information is suppressed by various speaker normalization techniques. Since biometric information is always present in EEG, similar methods need to be developed to scale BCIs across individuals. The subspace systems proposed in this paper gives a single vector representation of biometric information present in the signal. While building BCIs, the subject-specific vectors can also be used as features for normalizing variance across subjects.
 
 \subsubsection{Improving results on unseen tasks}
 \label{sec:fut:utasks}
 In Table~\ref{tab:6:mismatch-testing} of Section~\ref{sec:task_effect}, the performance of models were observed to drop significantly when very different conditions are used in training and testing. Similarly, for few tasks in Table~\ref{tab:left_out_task}, the result dropped considerably for few tasks unseen during training. In addition to this, in the supplementary material (Section S4), we present additional performance comparisons using baseline models for unseen tasks. The other baseline methods were also observed to have a similar drop in performances. Despite the drop in performance, it should be noted that these results are much better than a random classifier indicative of task-independent biometric signatures.  Benchmarking results on such hard inter-task testing and improving these results using better datasets and models are important future works. 
 
 \subsubsection{Improving results on unseen users}
 \label{sec:fut:uusers} 
  Table~\ref{tab:7:unseen_subject} of Section~\ref{sec:unseen} tests the ability of the proposed subspace embeddings to scale for unseen users during training. In the supplementary material (Section S5), we also report the results on unseen task for other compatible baseline models. For individuals unseen during training, the proposed model was shown to work despite a considerable drop in performance.  This experiment is inspired by speaker verification literature \cite{Snyder2018XVectorsRD,dehak2011front}, where embedding extracted from a subspace trained discriminatively using a large number of users ($\gg1000$) is the state-of-the-art for open-set-verification. The preliminary results given in Section~\ref{sec:unseen} show that the proposed models have the potential to be used in open-set verification by extracting embeddings from users unseen during model training. However, testing and improving open-set verification results using strong baselines and large datasets is an important future work for these models to be used for large-scale EEG biometric verification similar to speaker recognition. 
  
 \subsubsection{Better feature selection and channel sampling} 
 All results in the manuscript were reported using PSD features. Exploring better features for task-independent EEG biometric is an interesting future work. 
 
 In Section~\ref{sec:channel_effect}, various simple channel sub-sampling strategies are explored  using a $128$ channels dataset. Section S6 in the supplementary material reports other baseline results for all channel sub-sampling strategies discussed in Section~\ref{sec:channel_effect}. However, these channel sub-subsampling techniques are empirical. In this regard, in the supplementary  material (Section S7), we attempted a simple systematic way of sampling the channels by their performance; however, the obtained performances were not significantly greater than the results reported in Section~\ref{sec:channel_effect}. Using a dataset with $128$ channels, sampling channels systematically for every individual or a set of individuals using a compatible recognition system is also an interesting future direction for research.

\section{Conclusion}
\label{sec:Conclusion}
The paper builds upon state-of-the-art text-independent speaker recognition techniques, namely the $i$-vector and $x$-vector system, for recognizing individuals from multi-channel EEG. It proposes a novel system that combines adapted versions of $i$-vector and $x$-vector. The proposed methods are shown to give state-of-the-art results by testing against various baseline systems on two large datasets across tasks and sessions.  The proposed approach is shown to reliably encode person-specific signatures into a single vector using just four channels and a simple cosine similarity scoring. 

The proposed subspaces  are also used to offer empirical evidence for the presence of task-independent person-specific signatures in EEG using different tasks/conditions for training and testing at various levels. The results discussed in this paper question the need for the use of constrained elicitation protocols for EEG biometrics and suggest task-independent settings to be used as a baseline for studying task-dependent EEG biometrics.


%

\section*{Data Availability}
Dataset~1  is publicly available at \url{https://www.iitm.ac.in/donlab/cbr/eeg_person_id_dataset/}. The list of files used from Temple University database and the code will be shared on request.

\section*{Acknowledgment}
We thank the Centre for Computational Brain Research (CCBR), IIT Madras for enabling the collaboration between Sur Lab, Massachusetts Institute of Technology and Indian Institute of Technology Madras. M.S. holds the N.R. Narayanamurthy Visiting Chair Professorship in Computational Brain Research at IIT Madras.  Collaboration with University of Southern California (S.N) was supported by DST VAJRA.  Supported by the N.R. Narayanamurthy Chair endowment and DST VAJRA.

\ifCLASSOPTIONcaptionsoff
  \newpage
\fi



%

\bibliographystyle{IEEEbib}
{\scriptsize \bibliography{refs}}

\end{document}